\documentclass[a4paper,floatfix,nofootinbib,twocolumn]{revtex4-1}

\usepackage{epsfig}
\usepackage{amsmath,amsfonts,amssymb}
\usepackage{t1enc}

\usepackage{float}

\providecommand{\openone}{\leavevmode\hbox{\small1\kern-3.8pt\normalsize1}}

\usepackage{booktabs}
\usepackage[Q=yes,pverb-linebreak=no]{examplep}

\begin{document}

\title{Angular distributions in $t{\bar t}H (H\rightarrow b\bar{b})$ reconstructed events at the LHC}

\author{S.P. Amor dos Santos$^1$,  J.P. Araque$^2$, R. Cantrill$^3$, N.F. Castro$^{2,9}$, M.C.N. Fiolhais$^{1,4}$, \\
R. Frederix$^5$, R. Gon\c{c}alo$^{3}$, R. Martins$^{2}$, R. Santos$^{7,8}$, 
J. Silva$^6$, A. Onofre$^{2}$, H. Peixoto$^6$, A. Reigoto$^2$  \\[3mm]
{\footnotesize {\it 
$^1$ LIP, Departamento de F\'{\i}sica, Universidade de Coimbra, 3004-516 Coimbra, Portugal\\
$^2$ LIP, Departamento de F\'{\i}sica, Universidade do Minho, 4710-057 Braga, Portugal\\
$^3$ LIP, Av. Elias Garcia, 14-1, 1000-149 Lisboa, Portugal\\
$^4$ Department of Physics, City College of the City University of New York, \\ 
     160 Convent Avenue, New York 10031, NY, USA \\
$^5$ PH Department, TH Unit, CERN, CH-1211 Geneva 23, Switzerland \\
$^6$ Centro de F\'{\i}sica, Universidade do Minho, Campus de Gualtar, 4710-057 Braga, Portugal\\ 
$^7$ Instituto Superior de Engenharia de Lisboa - ISEL, 1959-007 Lisboa, Portugal \\
$^8$ Centro de F\'{\i}sica Te\'{o}rica e Computacional,
    Faculdade de Ci\^{e}ncias, Universidade de Lisboa, Campo Grande, Edif\'{\i}cio C8 1749-016 Lisboa, Portugal \\
$^9$ Departamento de F\'{\i}sica e Astronomia, Faculdade de Ci\^encias da \\[-1mm]
Universidade do Porto, Rua Campo Alegre 687, 4169 - 007 Porto, Portugal}}
}

\begin{abstract}
The associated production of a Higgs boson and a top-quark pair,
$t{\bar t} H$, in proton-proton collisions is addressed in this paper
for a center of mass energy of 13~TeV at the LHC. Dileptonic final
states of $t{\bar t}H$ events with two oppositely charged leptons and
four jets from the decays $t\rightarrow bW^+ \rightarrow b
\ell^+\nu_\ell$, $\bar{t}\rightarrow \bar{b}W^- \rightarrow \bar{b}
\ell^-\bar{\nu}_\ell$ and $h\rightarrow b\bar{b}$, are used. Signal
events, generated with MadGraph5\Q{_}aMC@NLO, are fully reconstructed
by applying a kinematic fit. New angular distributions of the decay
products as well as angular asymmetries are explored in order to
improve discrimination of $t{\bar t} H$ signal events over the
dominant irreducible background contribution, $t{\bar t}b{\bar b}$.
Even after the full kinematic fit reconstruction of the events, the
proposed angular distributions and asymmetries are still quite
different in the $t\bar{t}H$ signal and the dominant background
($t\bar{t}b\bar{b}$).
\end{abstract}


\maketitle

\section{Introduction}

On July 4th 2012 the ATLAS~\cite{:2012gk} and CMS~\cite{:2012gu}
collaborations announced the discovery of a scalar particle at
CERN's Large Hadron Collider (LHC). This new particle with a mass of
about $125$ GeV was later identified as the Higgs boson, responsible
for the generation of all particle masses through the mechanism of
spontaneous symmetry breaking~\cite{higgsmech}. So far, the measured
properties of the Higgs boson have shown a remarkable consistency with
those predicted by the Standard Model (SM) of particle
physics. Nevertheless, it is by now clear that the SM cannot explain
all the observed physical phenomena, as for instance it fails to
provide a candidate for dark matter or a means to explain the matter
anti-matter asymmetry in the Universe. However, as more data is being
accumulated and analysed at the LHC, it becomes increasingly clear that any new
physics theory has to resemble very much the SM at the electroweak
scale. In the first run, the ATLAS and CMS collaborations have studied
in great detail the main four Higgs production modes at the LHC~\cite{allexphiggs},
namely gluon fusion (including $b \bar b$ fusion), Vector Boson Fusion (VBF),
associated production ($VH$, with $V=W,Z$) and $t \bar t H$
production, with centre of mass energies of 7 and 8 TeV. For each production
mode several decay channels were considered and analysed in great detail.

The production of the Higgs boson in association with a top quark-anti-quark pair, $pp \to t \bar t H$~\cite{ttHtheory}, 
constitutes the only way (together with single top plus Higgs
which has an even smaller cross section) to directly probe the
top-quark Yukawa couplings.
Moreover, it is also contaminated by a
huge background coming mainly from $pp \to t \bar t+\textrm{
  jets}$. For this particular production process several decay
channels have been studied~\cite{Aad:2014lma,Aad:2015gra,Khachatryan:2015ila}.  
The very complex final states, together with the huge backgrounds to
the process, makes it the most difficult Higgs channel to study at the
LHC. Nevertheless, with just a few events, both collaborations have
reach a sensitivity down to about 2 -- 3 times the SM value which
constitutes a remarkable achievement.

The current studies~\cite{Aad:2014lma,Aad:2015gra,Khachatryan:2015ila}, 
use the kinematic information of the events to separate the signal from
the backgrounds. In this work we advocate the introduction of new variables 
that make use of the information from (lack of) spin correlations in the signal
and background processes~\cite{tth_spin,Artoisenet:2012st}: the top and anti-top quarks are natural spin
analysers of this process. We will show that part of the spin
information that is present in the matrix elements survives the parton
showering, detector simulation, event selection and event
reconstruction. These new variables could play an important role in
background discrimination, possibly leading to an improvement in
the precision of the measurement of the top-quark Yukawa
coupling. Even though we will consider only the irreducible $t{\bar t}b{\bar b}$
background, without a highly-optimized event-reconstruction method, we
will argue that our findings are also valid in a more general and
realistic case.

We should also note that the type of variables proposed in this work
can then be used to probe the CP nature of the top-Yukawa
coupling~\cite{ttHCPtest}. In many models like the CP-violating
two-Higgs doublet model~\cite{pheno2hdm} (the status of this model
after the LHC run 1 was recently presented in~\cite{Fontes:2015mea}), CP
violation appears explicitly in the Higgs sector via mixing of CP-even
and CP-odd states. The determination of the CP nature of the Higgs
boson and its interactions is of the utmost importance at the
LHC. Finally one should note that $t\bar{t}H$ production can be
studied at future linear colliders such as the ILC, which will lead to
a tremendous improvement in the precision of measurements of the
Yukawa couplings~\cite{ILC}.

\section{Signal and background generation at the LHC}
Given that the goal of this work is to study how well spin information
can be used to improve the current search strategies, we only consider
the signal and its dominant irreducible background. The signal ($t \bar t H$)
and background ($t \bar t b\bar b$) processes were generated, at
leading order (LO), using MadGraph5\Q{_}aMC@NLO~\cite{Alwall:2014hca}
with the NNPDF2.3 PDF sets~\cite{Ball:2012cx}. The full spin correlations information of
the $t\rightarrow bW^+ \rightarrow b \ell^+\nu_\ell$,
$\bar{t}\rightarrow \bar{b}W^- \rightarrow \bar{b}
\ell^-\bar{\nu}_\ell$ and $h\rightarrow b\bar{b}$ decays, with $l^\pm\in\{e^\pm,\mu^\pm\}$, was retained
by allowing MadSpin~\cite{Artoisenet:2012st} to perform the decay of
the heavy particles. Although other decay modes of the top quarks and Higgs boson could be considered,
in this paper we focus on the most challenging leptonic decay channel i.e., the dileptonic decay of the $t\bar{t}$ system 
together with a Higgs decaying to the dominant SM decay ($H\rightarrow b\bar{b}$). We argue the full 
kinematical reconstruction of the undetected neutrinos in such events, even if difficult, still preserves the angular 
distributions that could help in discriminating signal from irreducible backgrounds.
The events were generated for the LHC with a centre of mass
energy of 13~TeV with the default dynamic factorization and
renormalization scales, setting the masses of
the top quark and the SM Higgs boson to $172.5$~GeV and $125$~GeV,
respectively. We do not attempt to investigate possible departures
from the SM nature of the Higgs boson in this paper, assumed to be a
scalar particle ($CP=1$). The generated events were then passed to PYTHIA
6~\cite{Sjostrand:2006za} for shower and hadronization. In order to
obtain more realistic results, for example for differential cross
sections and efficiencies, we passed the generated events through
Delphes~\cite{deFavereau:2013fsa} to perform a fast detector
simulation of a general-purpose collider experiment at the LHC. We
used the ATLAS default card for the simulation and performed the analysis of the generated 
and simulated events with MadAnalysis 5~\cite{Conte:2012fm} in the expert mode~\cite{Conte:2014zja} .

The full kinematical reconstruction of $t\bar{t}H$ events is very
challenging in the dileptonic decays of the $t\bar{t}$ system, 
since both undetected neutrinos need to be reconstructed. In this paper, we
explore the advantages of fully reconstructing the $t\bar{t}H$ system
in the dileptonic topology, by applying a kinematic fit to the
events using mass constraints and energy-momentum conservation. 
Events, after detector simulation, are accepted if they had at least
four reconstructed jets and two charged leptons with transverse momentum $p_T\ge20$~GeV
and pseudo-rapidity $\eta\le2.5$. No cuts are applied to the events transverse missing
energy ($\slash\kern-.6emE_{T}$).

In the following we will normalise the distributions of signal and
background to equal area, irrespective of their (fiducial) cross
sections and efficiencies: for our goal it suffices to show which
observables are sensitive to the difference in spin information in the
signal and background events and how well this information can be
retained in a realistically reconstructed event. For the same reason
we also abstain from performing a careful analysis of the
uncertainties in the event generation as well as the inclusion of
next-to-leading order corrections in the strong coupling.

\section{Reconstruction of dilepton $t\bar{t}H$ events after detector simulation}
As previously stated we will perform $t\bar tH$ event reconstruction in final
states with two charged leptons and at least four jets, after Delphes
simulation. We do not attempt to tag the flavour of jets from the
hadronization of $b$ quarks, i.e., we do not use any $b$-tagging tool to
help with the identification of the heavy flavour component of jets;
a task left outside the scope of this paper. The full kinematical
reconstruction requires the knowledge of the jet and charged lepton momenta,
together with the transverse missing energy. We use the $W$ and the top quark 
masses as constraints. The Higgs boson mass ($m_H=125$~GeV) is used to maximise the 
probability of the best combination of two jets chosen among the ones which were not utilised 
in the $t\bar{t}$ system kinematical reconstruction.
The transverse missing energy is re-fitted to improve the resolution of the
experimental measurement. After applying the constraints, six unknowns
need to be fully reconstructed in the dileptonic $t\bar tH$ events,
which are the 3-momenta of the two neutrinos present in the events. To
find a kinematic solution we assume the neutrinos are responsible for
the missing transverse energy, i.e.,
\begin{eqnarray}
\label{equ:a0}
\ensuremath{p_{x}^{\nu} + p_{x}^{\bar{\nu}} &=& \slash\kern-.6emE_{x}}, \\
\label{equ:a1}
\ensuremath{p_{y}^{\nu} + p_{y}^{\bar{\nu}} &=& \slash\kern-.6emE_{y}}.
\label{equ:a2}
\end{eqnarray}
In addition we apply the following mass constraints to the $t\bar{t}$ system of the events, 
\begin{eqnarray}
\ensuremath{( p_{\ell +} + p_{\nu} )^{2} &=& m_{W}^{2}}, \\
\label{equ:a3}
\ensuremath{( p_{\ell -} + p_{\bar{\nu}} )^{2} &=& m_{W}^{2}}, \\
\label{equ:a4}
\ensuremath{( p_{W^+} + p_{b} )^{2} &=& m_{t}^{2}}, \\
\label{equ:a5}
\ensuremath{( p_{W^-} + p_{\bar{b}} )^{2} &=& m_{t}^{2}}.
\label{equ:a6}
\end{eqnarray}
While $\ensuremath{\slash\kern-.6emE_{x}$ and $\slash\kern-.6emE_{y}}$
represent the $x$ and $y$ components of the transverse missing energy,
$\ensuremath{p_{\ell +}$ and $p_{\ell -}}$ ($\ensuremath{p_{b}$ and
  $p_{\bar{b}}}$) correspond to the two lepton (two $b$-jets) four
momenta, respectively, from the $t$ ($\bar{t}$)
decays. $\ensuremath{m_{W}$ and $m_{t}}$ are the $W$-boson and top
quark masses, respectively. The mass of the $W$-boson was set to $80.4$~GeV.

\begin{figure*}
\begin{center}
\begin{tabular}{ccc}
\epsfig{file=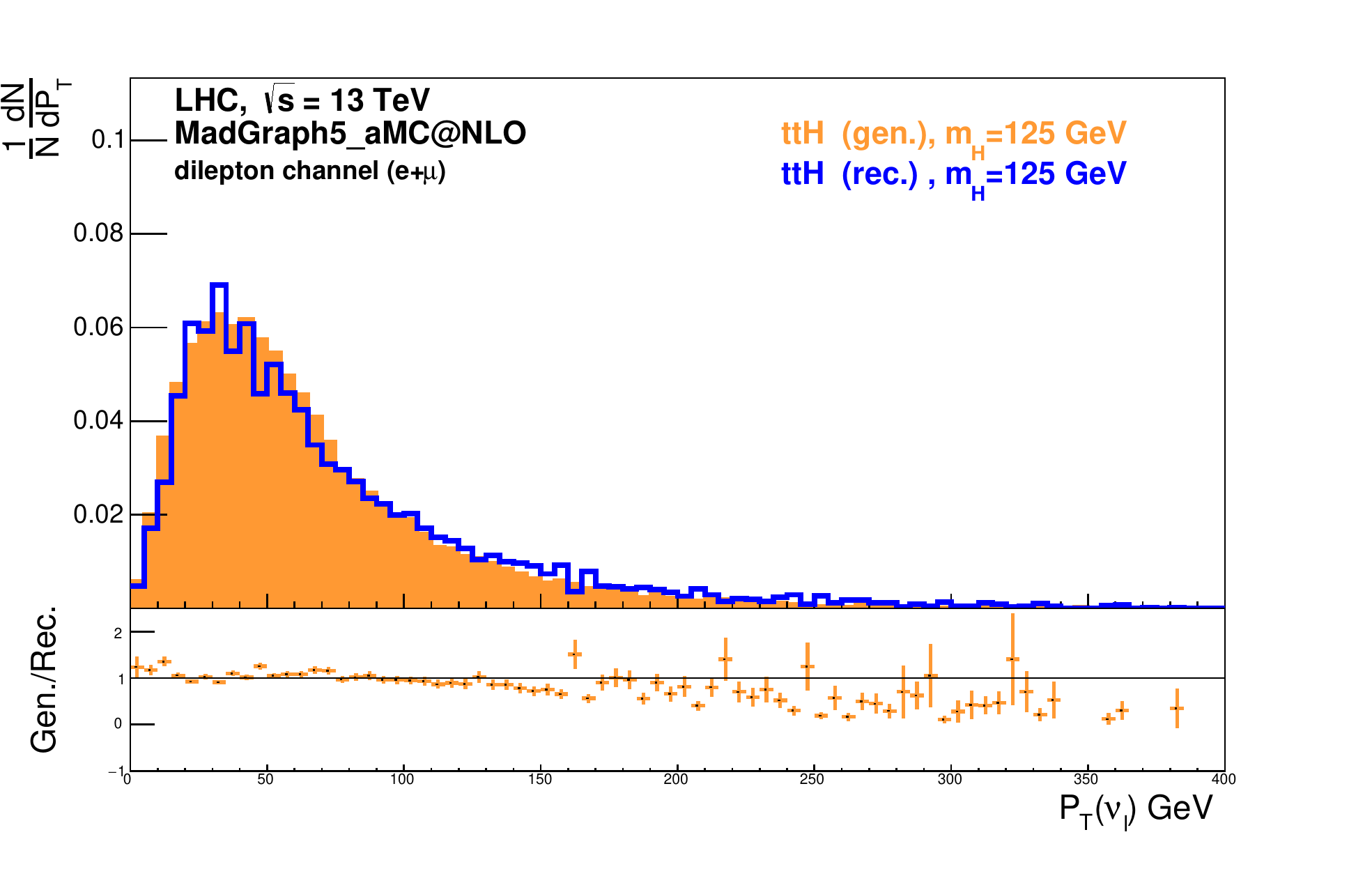,height=5.5cm,width=8.5cm,clip=} & \quad &
\epsfig{file=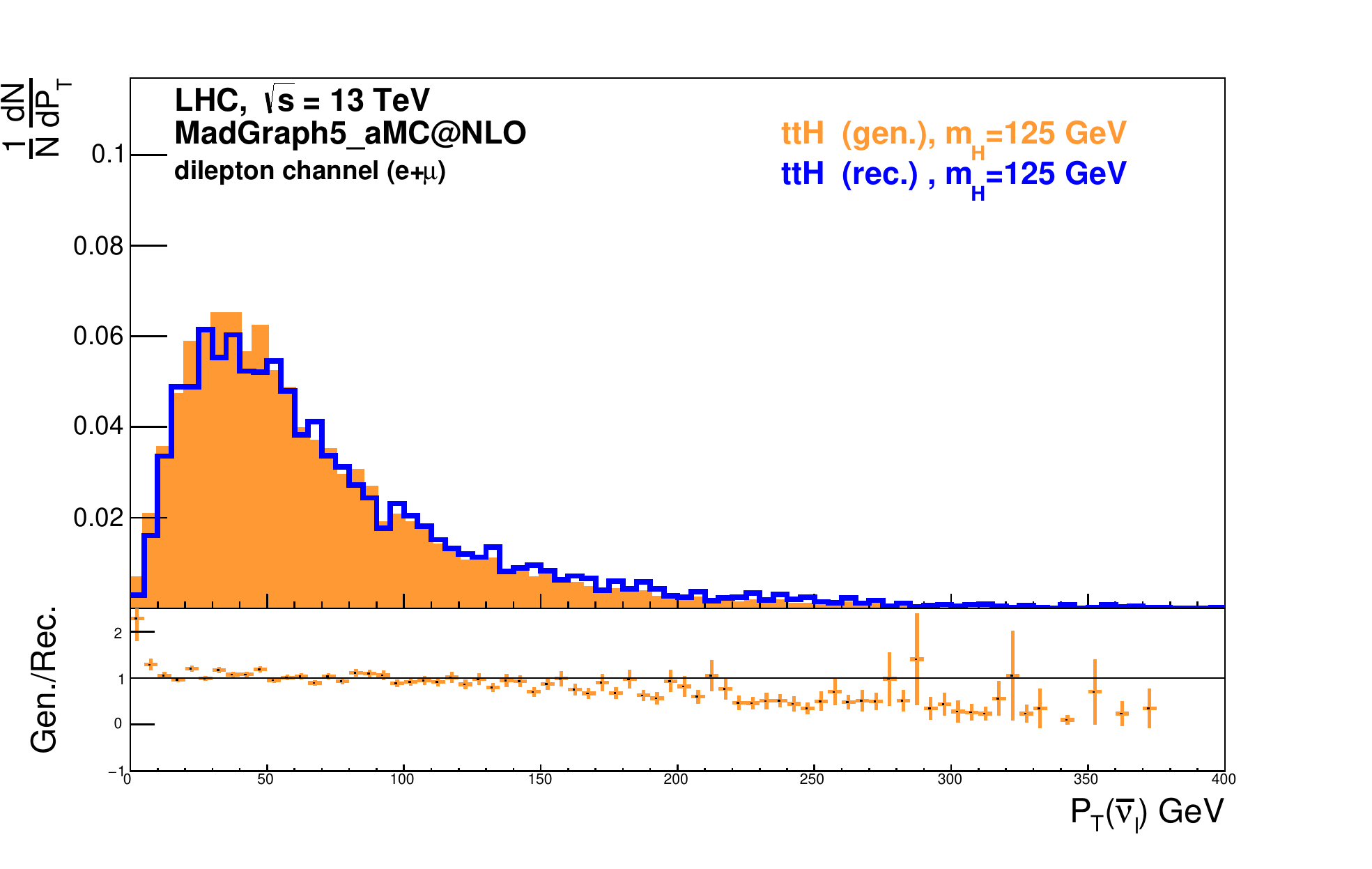,height=5.5cm,width=8.5cm,clip=}
\end{tabular}
\caption{Neutrino (left) and antineutrino (right) $p_{T}$ distributions. The generated distribution (shadowed region) is compared with the kinematical fit reconstruction with truth match (full line) distribution.}
\label{fig:Genrec_Rec1}
\end{center}
\end{figure*}

\begin{figure*}
\begin{center}
\begin{tabular}{ccc}
\epsfig{file=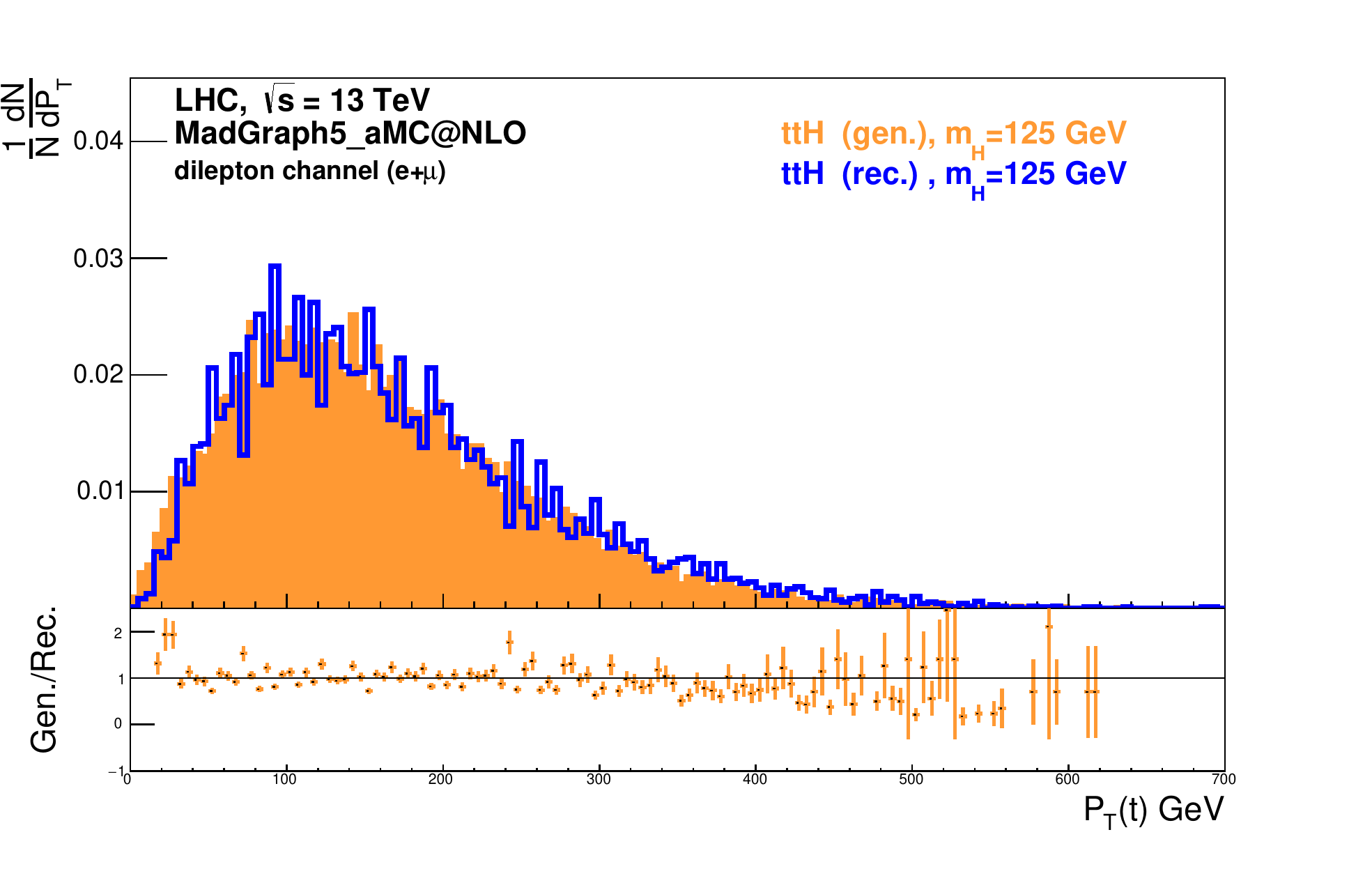,height=5.5cm,width=8.5cm,clip=} & \quad &
\epsfig{file=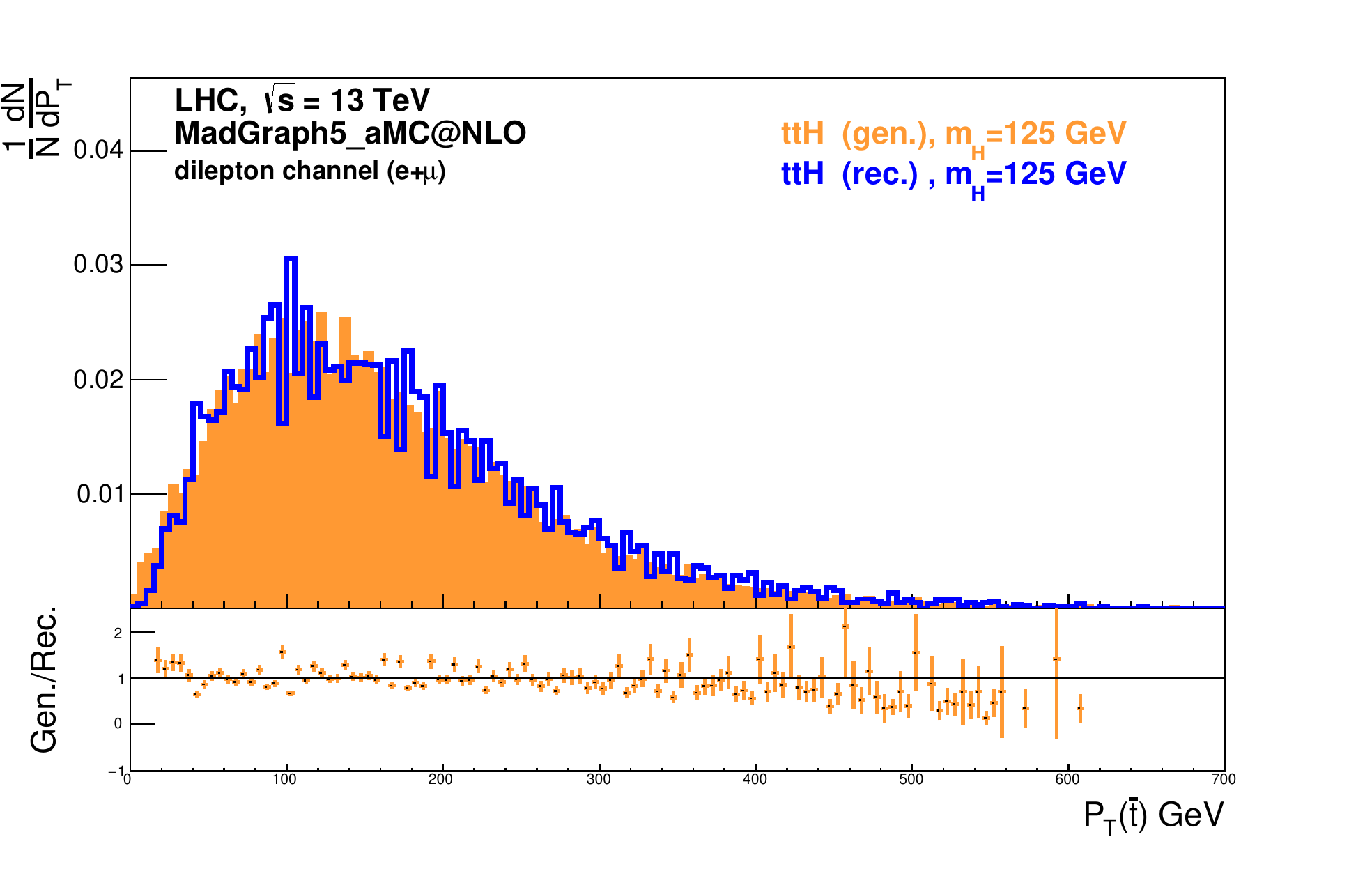,height=5.5cm,width=8.5cm,clip=}
\end{tabular}
\caption{Same as in figure~\protect\ref{fig:Genrec_Rec1}, but for the top (left) and anti-top quarks (right) $p_{T}$ distributions.}
\label{fig:Genrec_Rec2}
\end{center}
\end{figure*}

\begin{figure*}
\begin{center}
\begin{tabular}{ccc}
\epsfig{file=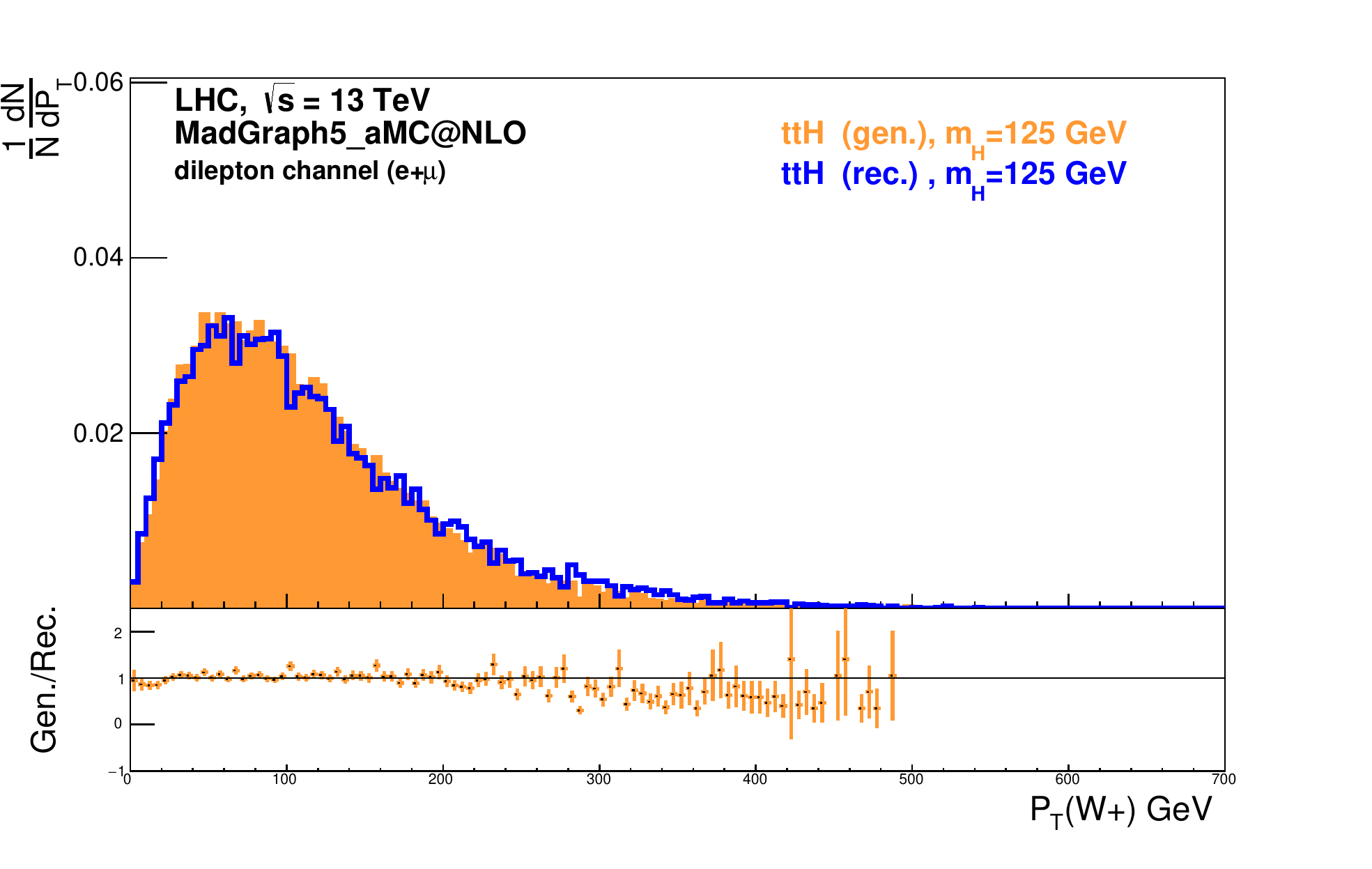,height=5.5cm,width=8.5cm,clip=} & \quad &
\epsfig{file=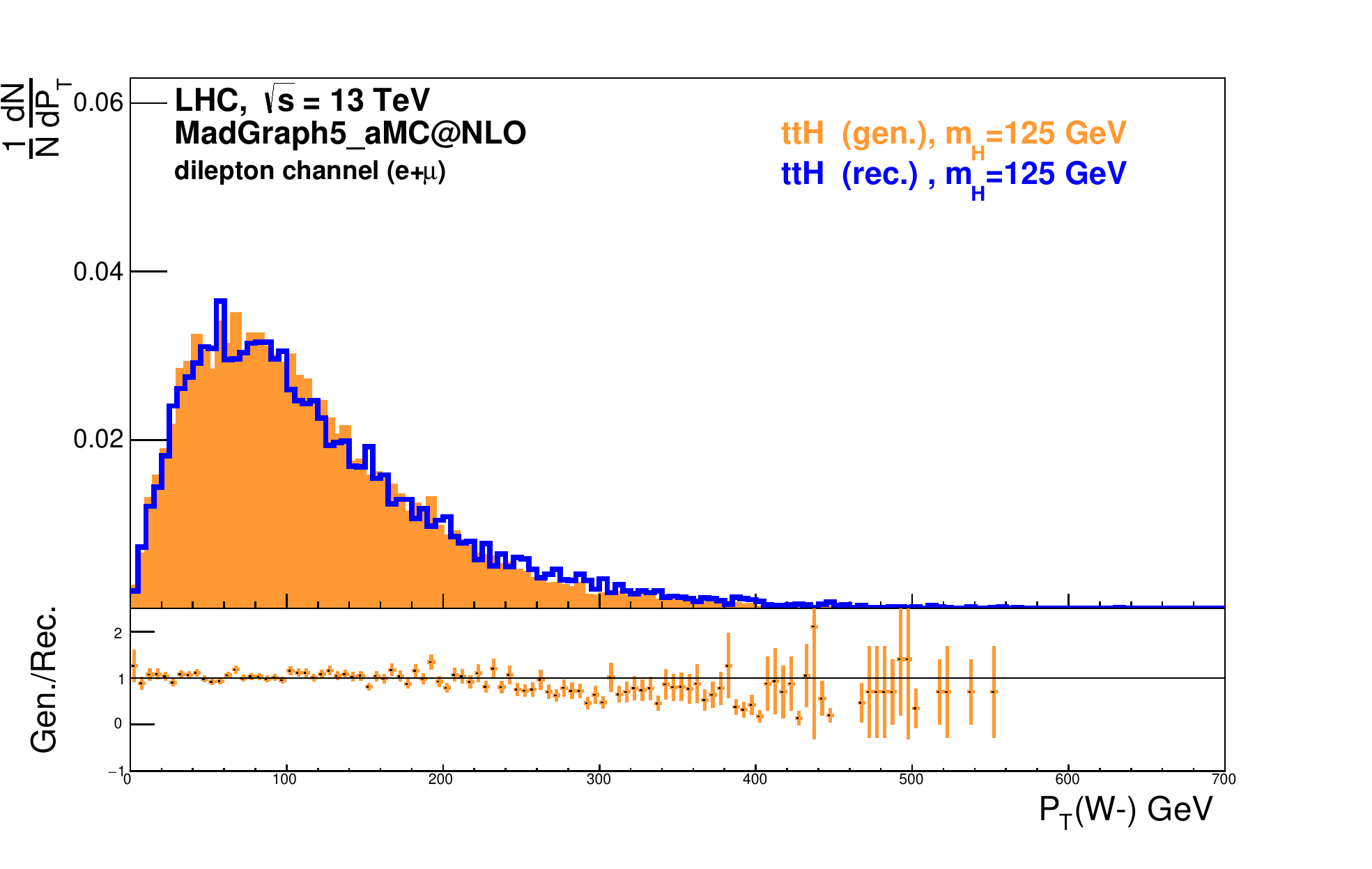,height=5.5cm,width=8.5cm,clip=}
\end{tabular}
\caption{Same as in figure.~\protect\ref{fig:Genrec_Rec1}, but for the $W^+$ (left) and $W^-$ (right) $p_{T}$ distributions.}
\label{fig:Genrec_Rec3}
\end{center}
\end{figure*}

We study the performance of the reconstruction with respect to the
generated parton-level Monte Carlo information. To make sure the kinematical
reconstruction produces sensible results, the
reconstruction is first applied to truth-matched objects, i.e., jets and
leptons which are matched to their parton-level generated quarks and
charged leptons, using a $\Delta{R}$ criterion (the minimum distance in the pseudorapidity-azimuthal angle plane, 
$\Delta{R}$, between the reconstructed jet or lepton and the parton-level quark or charged lepton, ensures the matching). Even though we use a
rather simple kinematical reconstruction method the efficiency using
truth-matched objects is 62\%. In figure~\ref{fig:Genrec_Rec1} the
neutrino (left) and antineutrino (right) $p_{T}$ from signal events
are shown. The generated distributions (filled histograms) are compared
with the truth-match reconstructed ones (solid lines). In the bottom plot, the ratio between the two distributions is
shown. Although a slight slope is visible in the ratio plot, more significant at high $p_T$ due to radiation effects not 
explicitly corrected for at the moment, good agreement between the reconstructed distributions is
observed with respect to the parton-level neutrino distributions, making clear that the full kinematical reconstruction of $t\bar{t}H$ events is possible. 
In figures~\ref{fig:Genrec_Rec2} and~\ref{fig:Genrec_Rec3} the $p_T$
distributions of the $t$ ($\bar{t}$) quarks and $W^+$ ($W^-$) bosons are
shown, respectively, for the $t\bar{t}H$ events.  Once again we see a similar
behaviour as observed for the neutrino $p_T$ distributions, i.e.,
good agreement between the reconstructed kinematic distributions and
the corresponding ones at parton level, in spite of the slight slope for higher $p_T$ values, 
in the ratio plot. Although the kinematical fit can correct, to a large extend, the effects of radiation, at high values the differences between reconstructed and
generated distributions may require an additional correction. Even though this would not difficult to implement, we have decide not to apply it here once
it may depend on the exact experimental environment conditions and does not contribute significantly to the main discussion of the
paper.

\begin{figure*}
\begin{center}
\vspace*{3.0cm}
\begin{tabular}{ccc}
\epsfig{file=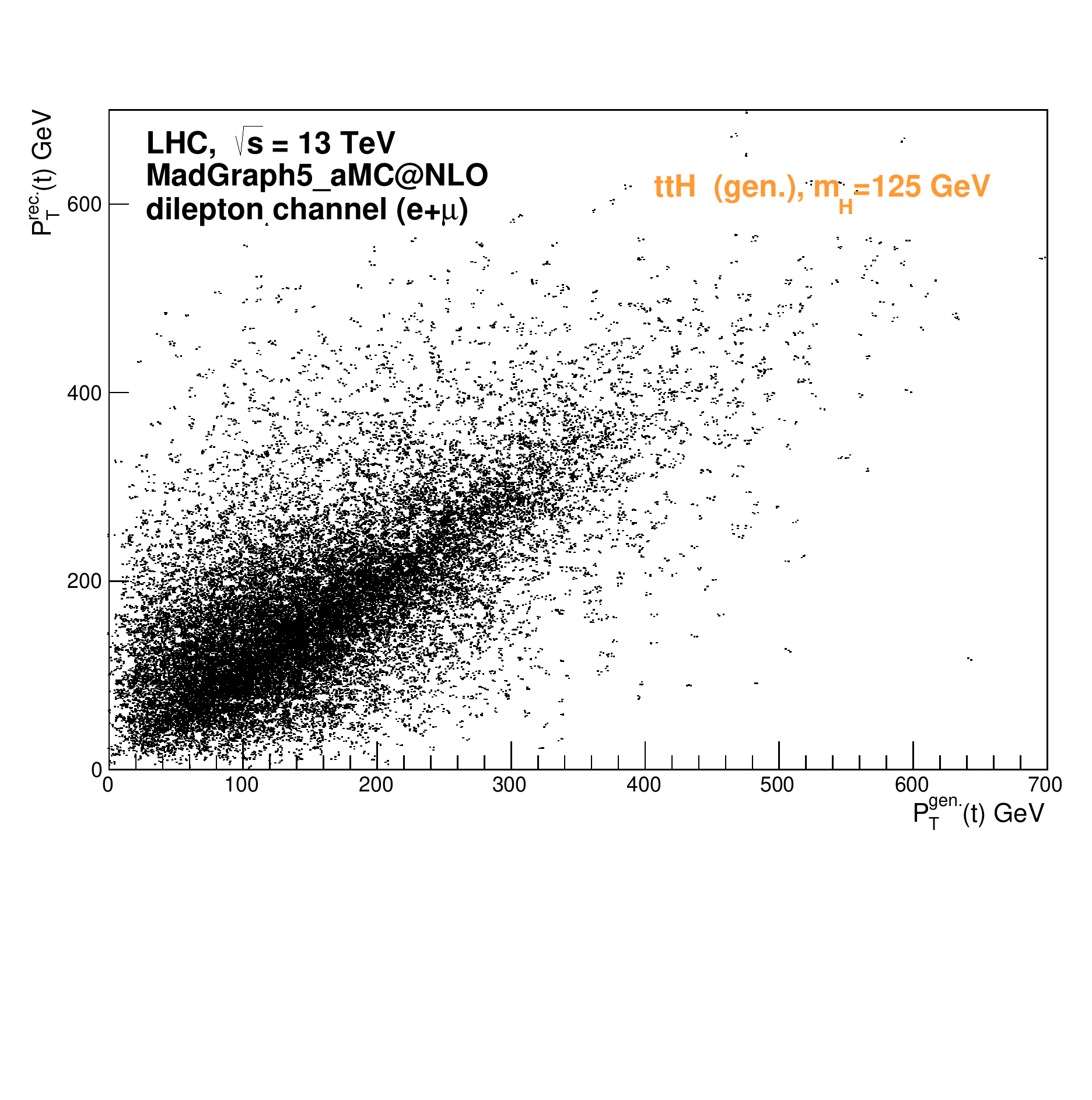,height=8.5cm,width=8.5cm,clip=} & \quad &
\epsfig{file=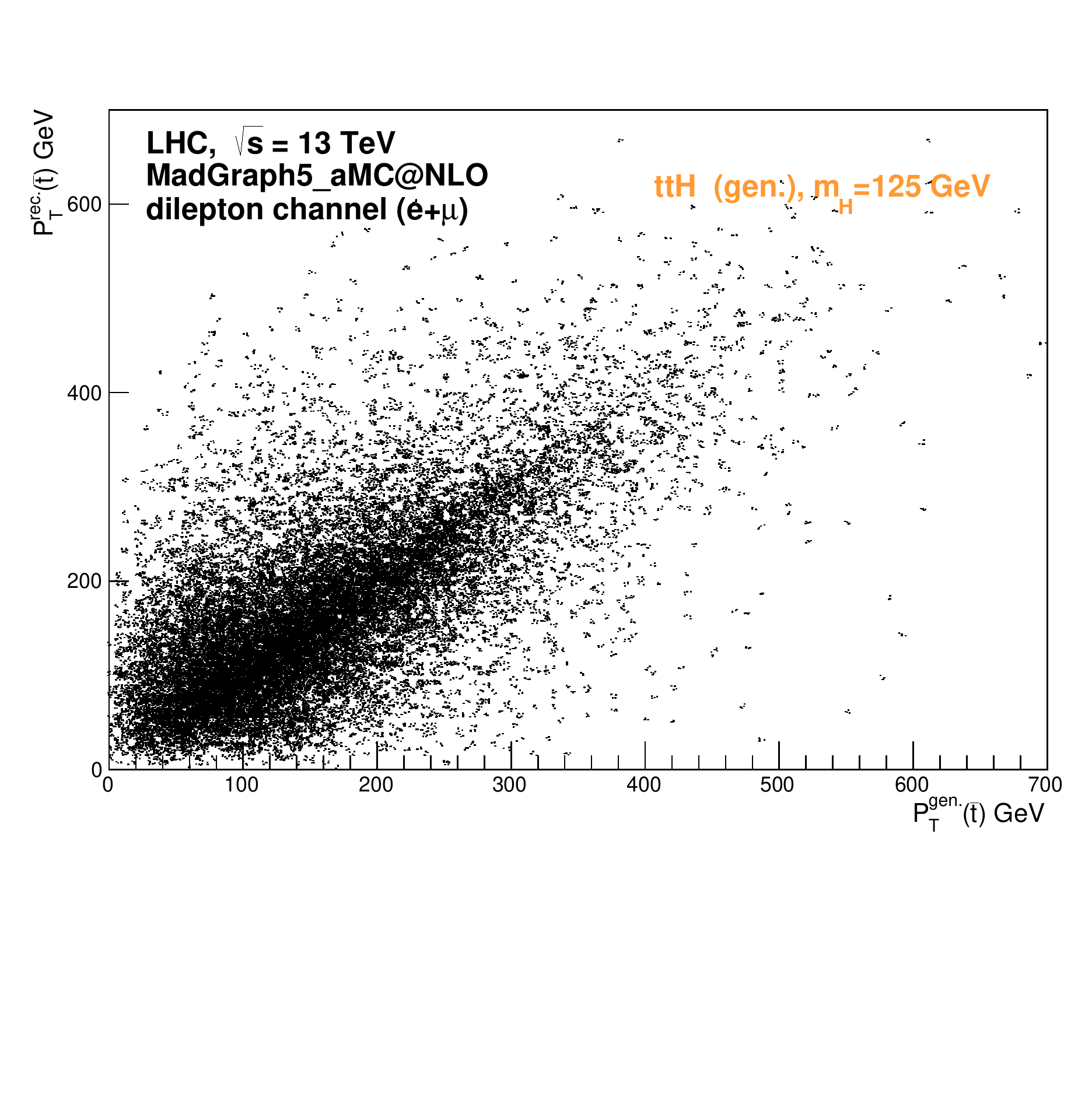,height=8.5cm,width=8.5cm,clip=}
\end{tabular}
\vspace*{-2.0cm}
\caption{Reconstructed top (left) and anti-top (right) quark $p_T$ using the kinematical fit (without truth match) as a function of the $p_T$ at parton level.}
\label{fig:Genrec_Exp2}
\end{center}
\end{figure*}

\begin{figure*}
\begin{center}
\vspace*{-2.cm}
\begin{tabular}{ccc}
\epsfig{file=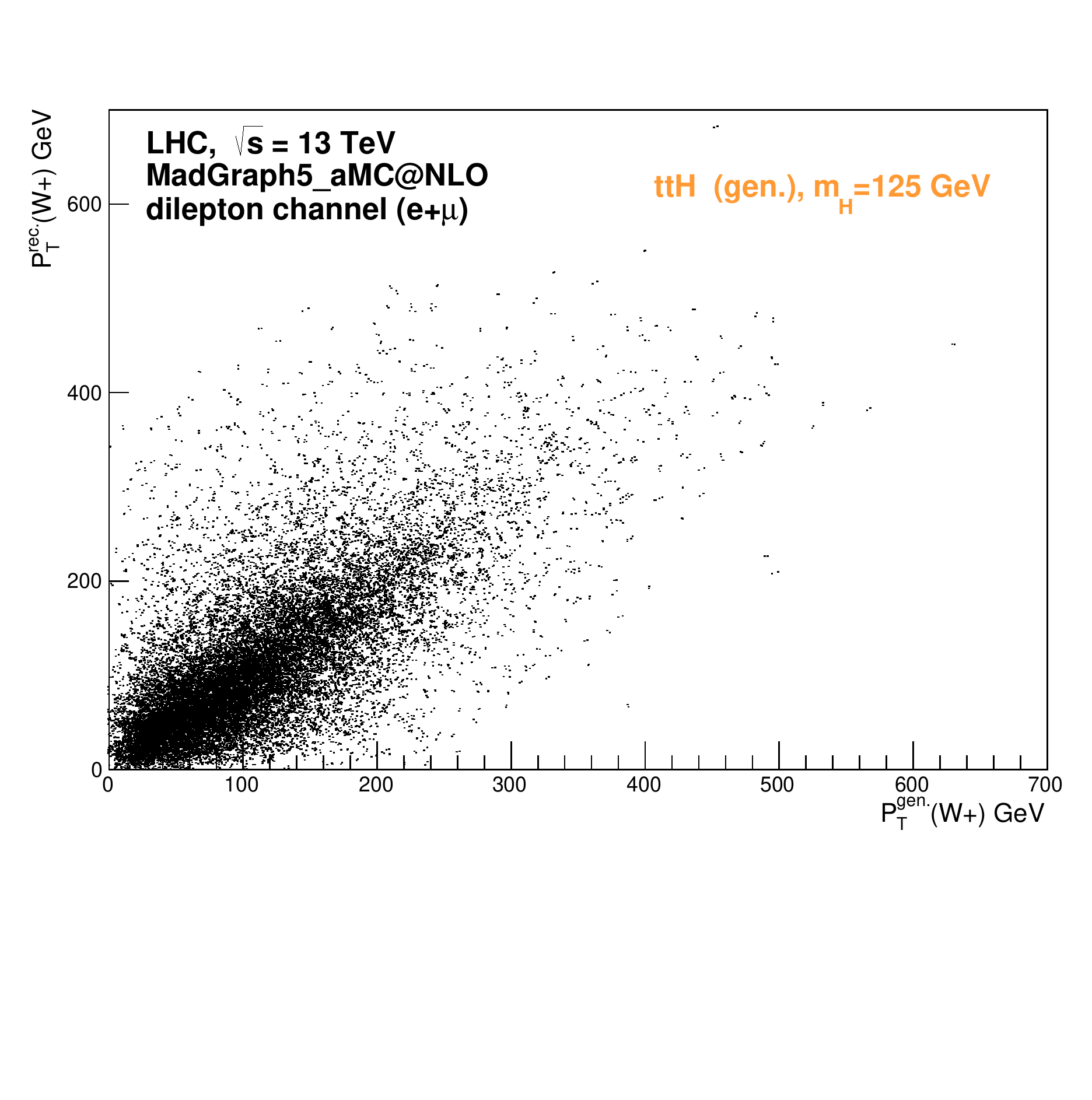,height=8.5cm,width=8.5cm,clip=} & \quad &
\epsfig{file=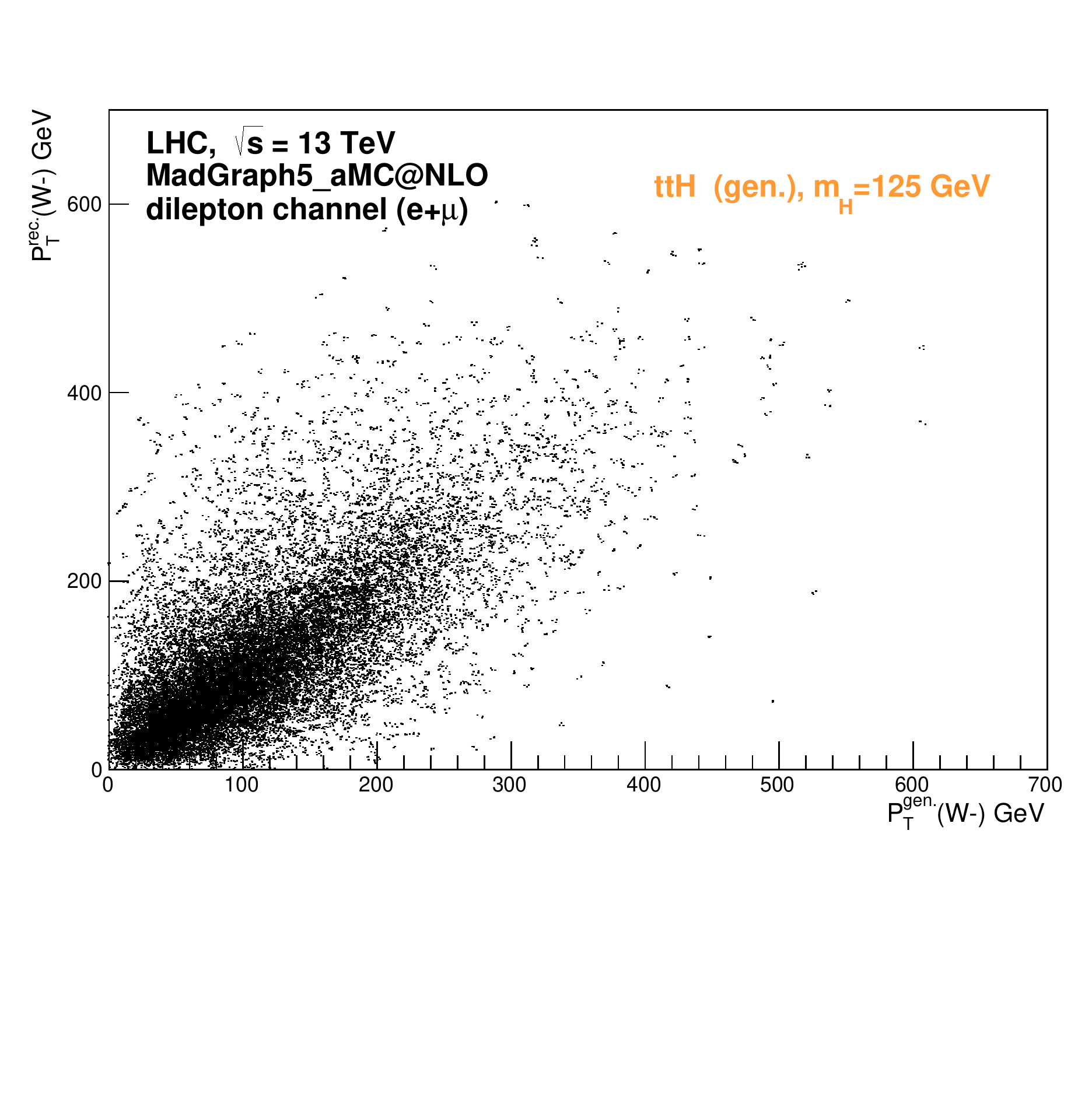,height=8.5cm,width=8.5cm,clip=}
\end{tabular}
\vspace*{-2.0cm}
\caption{Same as in figure.~\protect\ref{fig:Genrec_Exp2}, but for the $W^+$ (left) and $W^-$ (right) $p_{T}$ distributions.}
\label{fig:Genrec_Exp3}
\end{center}
\end{figure*}

In a second step, the truth match condition is dropped, bringing the
analysis closer to what can be done at collider
experiments. For this particular case we perform all possible
combinations of reconstructed jets and charged leptons (after detector
simulation) in order to reconstruct the top and anti-top quarks, together with the Higgs boson. For the $t\bar{t}$ system 
reconstruction we used the same procedure based on equations eq.~\eqref{equ:a0}-\eqref{equ:a6}.
We calculate the probability $P_{t\bar{t}}$ that the event is compatible with the equations, using probability density functions
for the neutrino and anti-neutrino $p_T$ distributions, the top and anti-top quarks mass distributions as well as the $W^+$ and $W^-$ bosons mass distributions, obtained at parton level.
To identify the two-jet combination, among the ones not used in the $t\bar{t}$ reconstruction, that best matches the jets from the Higgs boson decay, we associate to each 
combination, a weight $P_H$,
\begin{eqnarray}
\ensuremath{P_H & = & 1/|  \sqrt{(p_{i} + p_{j} )^{2}} - m_{H}|},
\label{equ:a7}
\end{eqnarray}
related to how close the  Higgs boson mass ($m_H=125$~GeV) is to the invariant mass of each particular jet-pair combination. The solution with highest
$P_{t\bar{t}} \times P_H$ is chosen as the right one for the full kinematical reconstruction of the events. This fixes completely the 
assignment of jets and charged leptons to their parent $t$, $\bar{t}$ quarks and Higgs boson. 
Due to the increase in the number of possible combinations which can
satisfy eq.~\eqref{equ:a0}-\eqref{equ:a7}, 88\% of all events are reconstructed by the kinematic
fit. This will obviously lead to an increase of the combinatorial background but, as we will see later, the kinematics are in most cases 
distinct from the right combinations.
%
%
In figures~\ref{fig:Genrec_Exp2} and ~\ref{fig:Genrec_Exp3}
we show the $p_T$ distributions of the top quarks and $W$ bosons,
respectively. The kinematically reconstructed $p_T$, with no jets and leptons truth match, is compared with the
parton-level distribution. We see a good correlation between the
kinematically reconstructed distributions with respect to the parton-level ones, thus ensuring that the reconstruction works fairly well. We
did not attempt to further optimise the event reconstruction because, again,
the main goal here is to show that a reconstruction is
possible with a reasonable efficiency.

\section{Angular distributions}

We will focus on angular distributions in fully reconstructed $t\bar{t}H$ events involving three-dimensional angles between the decay products of the $t\bar{t}H$ dileptonic final states. Following the full reconstruction of events, we define two reference frames:
\begin{itemize}
\item Frame 1: the full $t\bar{t}H$ centre-of-mass system, built by using the laboratory four-momenta and,
\item Frame 2: the $\bar{t}H$ centre-of-mass system recoiling against the $t$ quark, in the $t\bar{t}H$ system (i.e., in Frame 1 as defined above). 
\end{itemize}
For the generated distributions (with and without the $p_T$ and $\eta$ cuts applied in the event selection), we
use the parton-level four-momenta of all relevant objects. For the
reconstructed distributions (with and without truth match), we use the
four-momenta obtained after applying the kinematic fit
reconstruction. We define the angle between the Higgs momentum
direction (in the $\bar{t}H$ centre-of-mass) with respect to the
$\bar{t}H$ direction (in the $t\bar{t}H$ system) as $\theta^{\bar{t}
  H}_{H}$ and the angle of the $Y$ top quarks or Higgs decay products
($W^+$,$W^-$,$\ell^+$,$\ell^-$, $b$ and $\bar{b}$ jets) momentum (in
the Higgs centre-of-mass system) with respect to the Higgs direction
(in the $\bar{t}H$ system) as $\theta^{H}_{Y}$. We should stress the fact
that, when boosting $Y$ to the centre of mass of the Higgs boson, the
laboratory four-momenta were used (in a direct,
rotation-free boost).

\begin{figure*}
\begin{center}
\begin{tabular}{ccc}
\epsfig{file=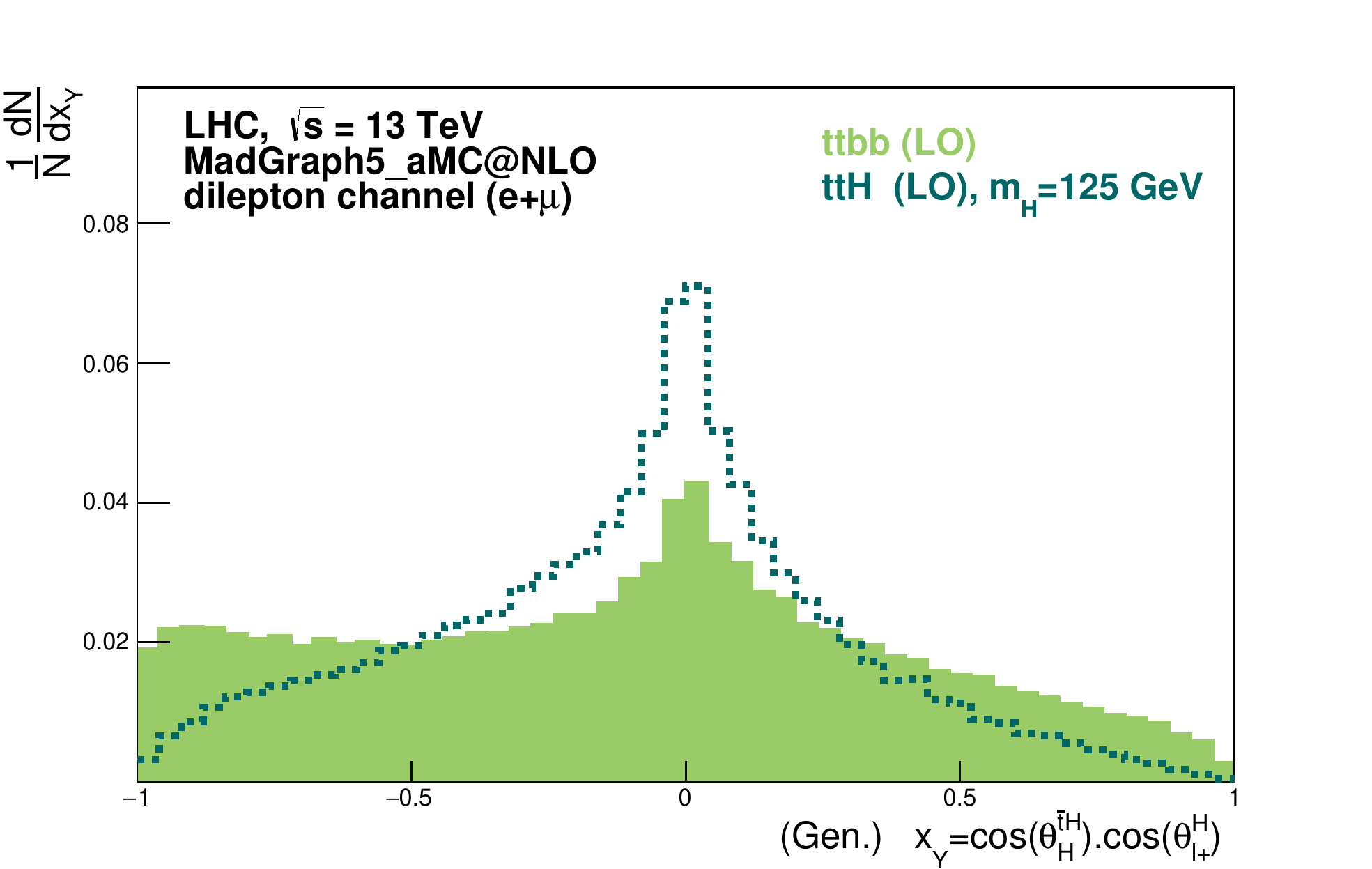,height=5.5cm,clip=} & \quad &
\epsfig{file=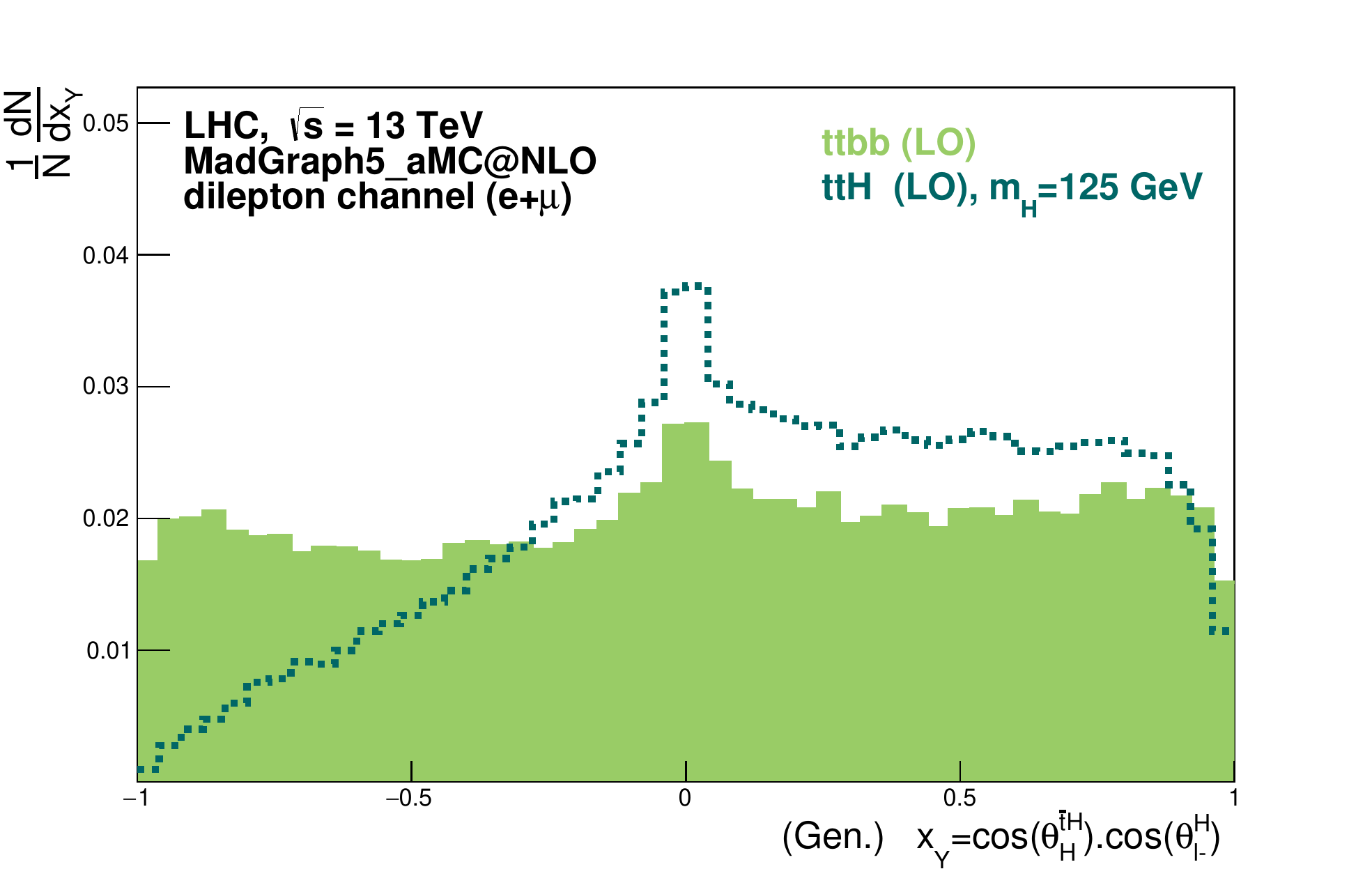,height=5.5cm,clip=}
\end{tabular}
\caption{Generated product of the cosine of the angle between the Higgs momentum direction (in the $\bar{t}H$ centre-of-mass) with respect to the $\bar{t}H$ direction (in the $t\bar{t}H$ system),  and the cosine of the angle of the $\ell^+$(left) and $\ell^-$(right) momentum (in the Higgs centre-of-mass system) with respect to the Higgs direction (in the $\bar{t}H$ system).}
\label{fig:GenNCAng1}
\end{center}
\end{figure*}

\begin{figure*}
\begin{center}
\begin{tabular}{ccc}
\epsfig{file=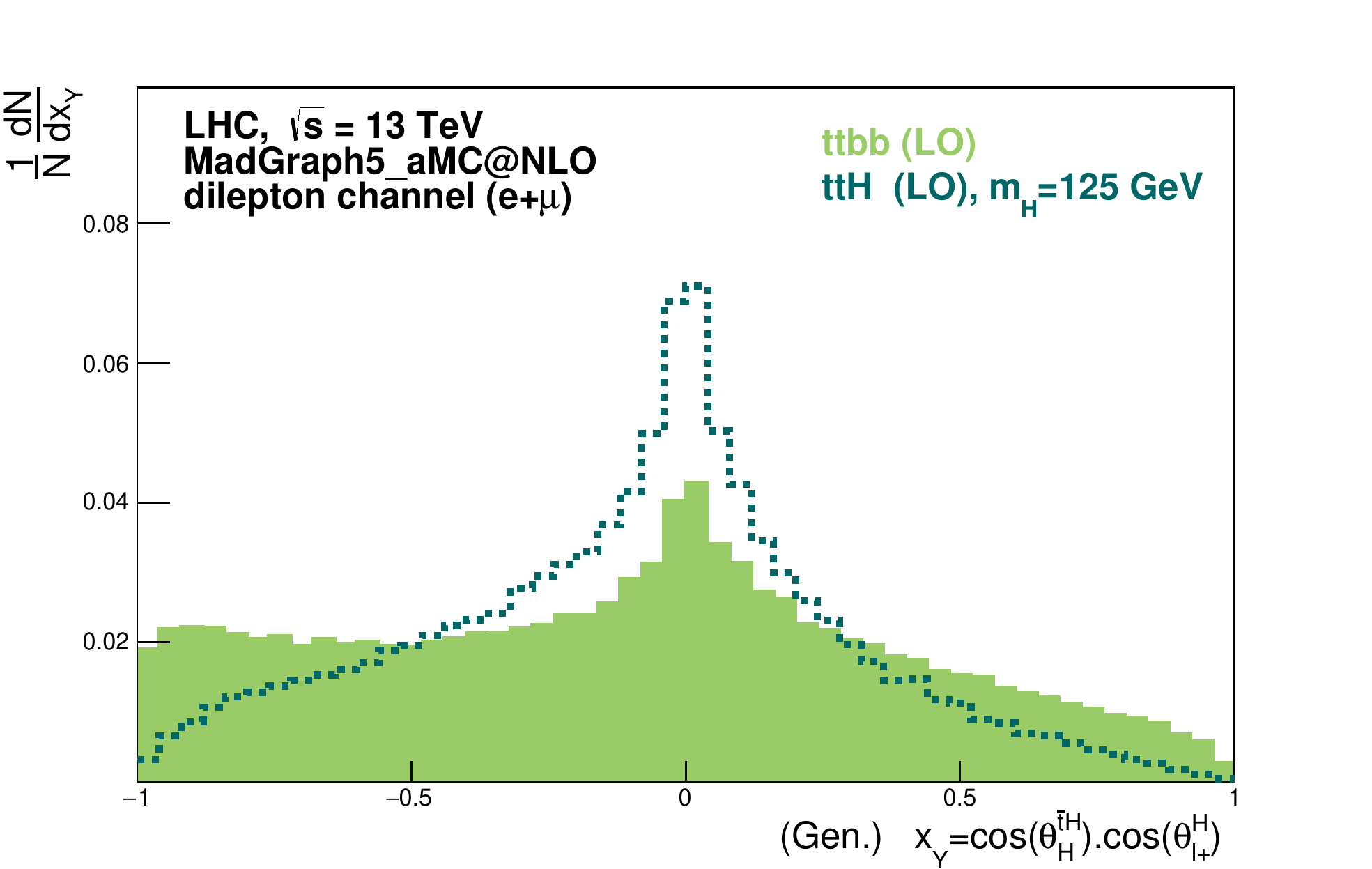,height=5.5cm,clip=} & \quad &
\epsfig{file=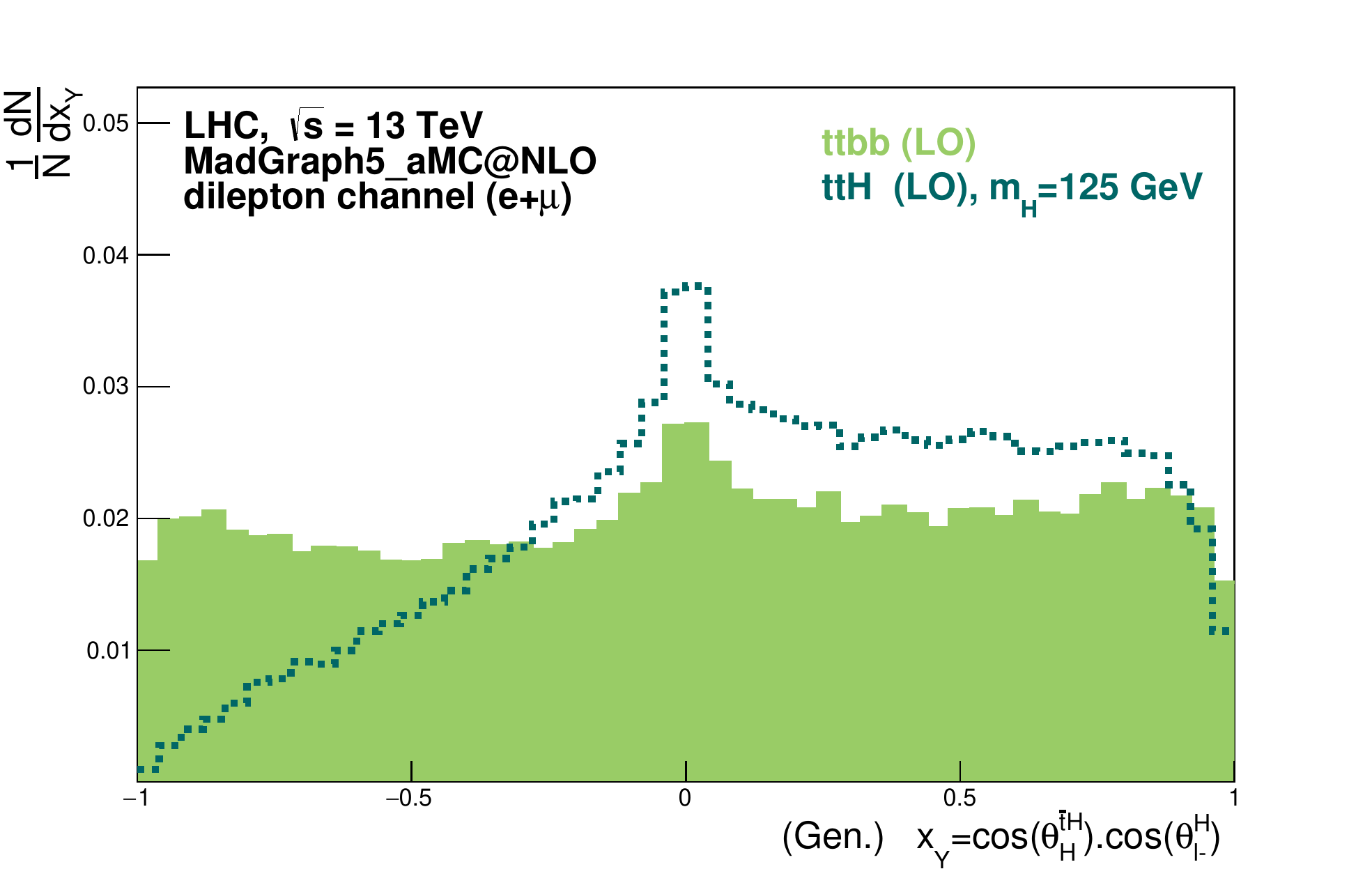,height=5.5cm,clip=}
\end{tabular}
\caption{Same as in figure.~\protect\ref{fig:GenNCAng1}, but after applying the acceptance cuts.}
\label{fig:GenWCAng1}
\end{center}
\end{figure*}

\begin{figure*}
\begin{center}
\begin{tabular}{ccc}
\epsfig{file=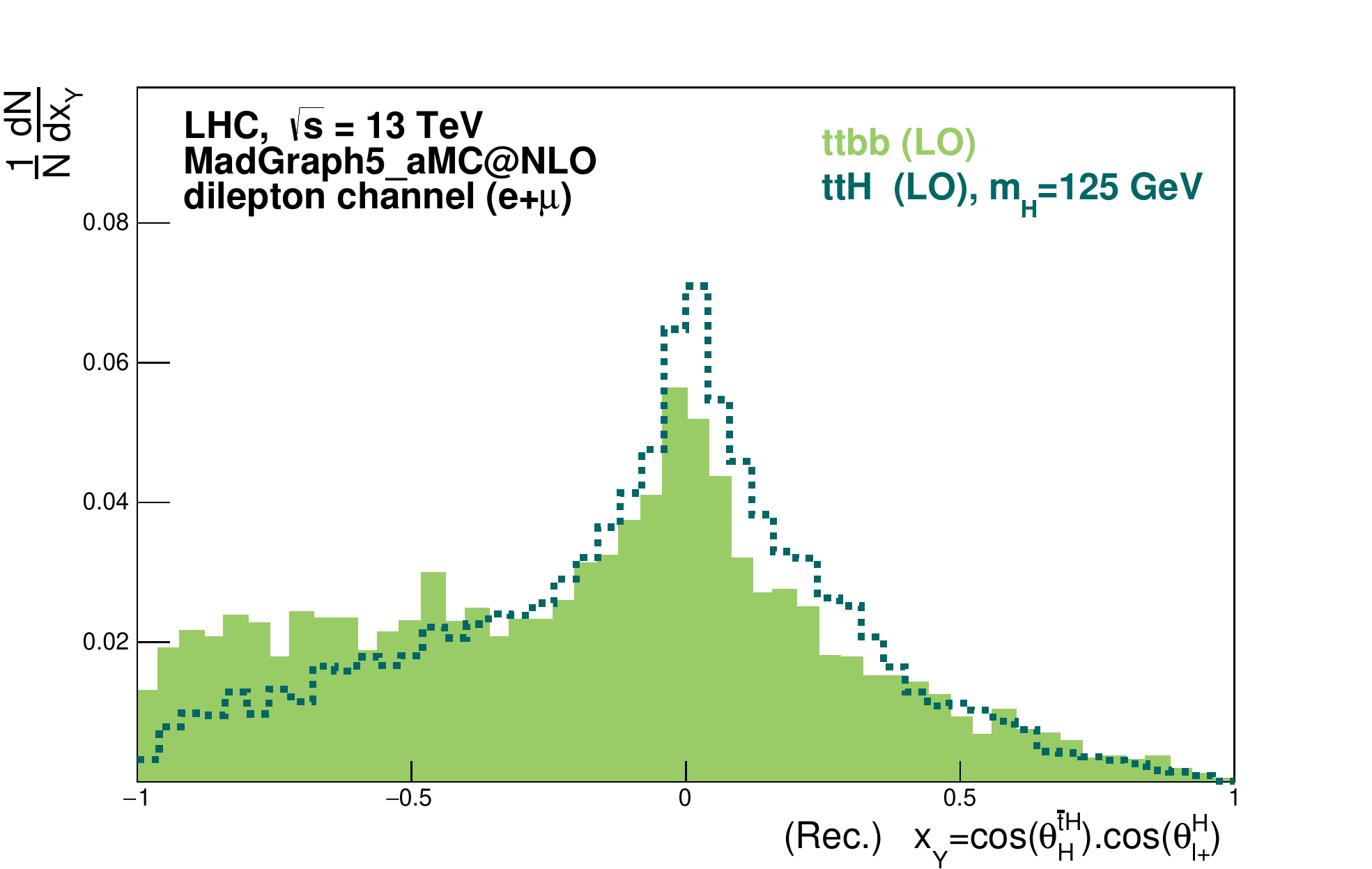,height=5.5cm,clip=} & \quad &
\epsfig{file=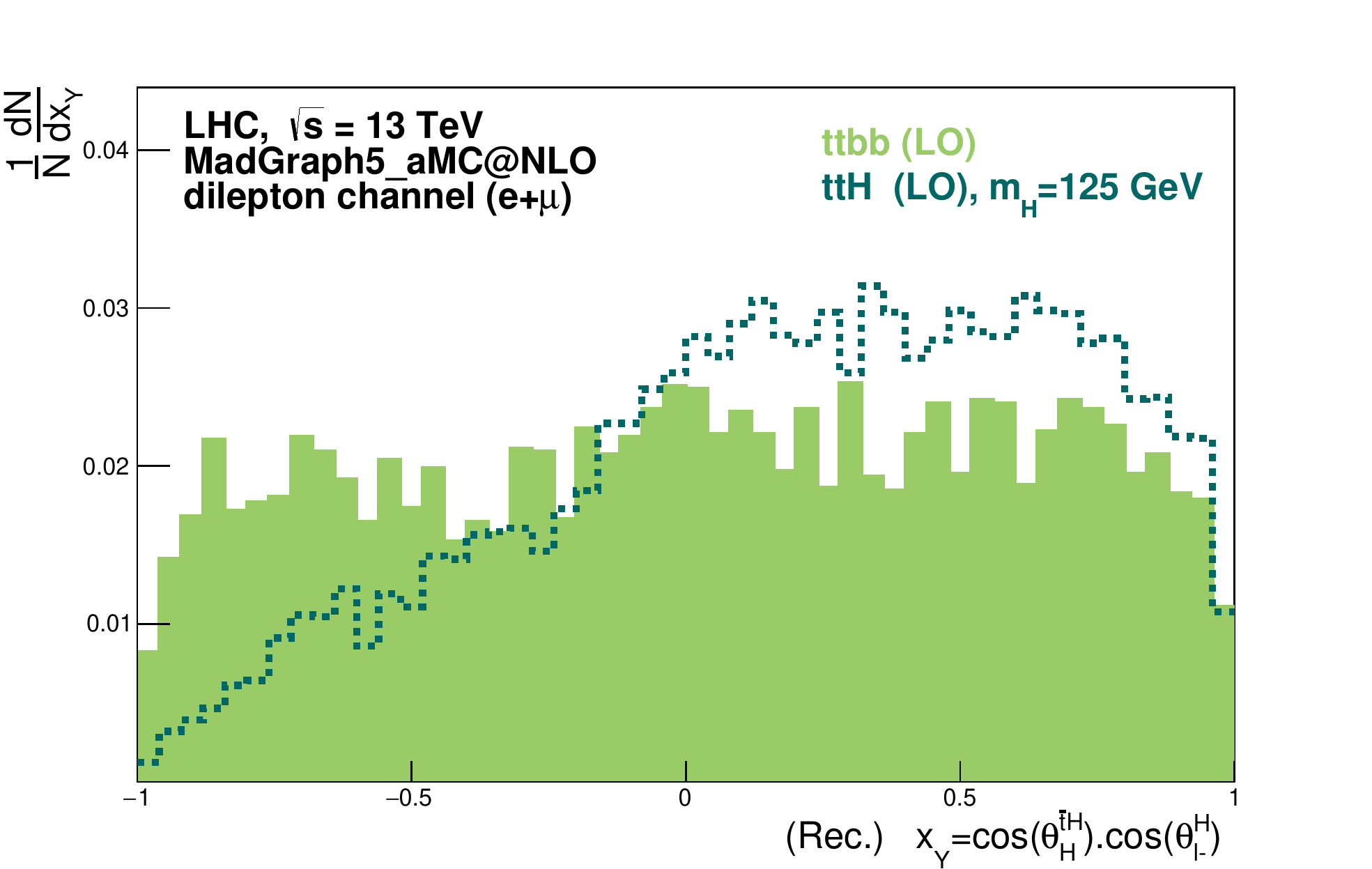,height=5.5cm,clip=}
\end{tabular}
\caption{Same as in figure.~\protect\ref{fig:GenWCAng1}, but using reconstructed objects with truth match.}
\label{fig:RecAng1}
\end{center}
\end{figure*}

\begin{figure*}
\begin{center}
\begin{tabular}{ccc}
\epsfig{file=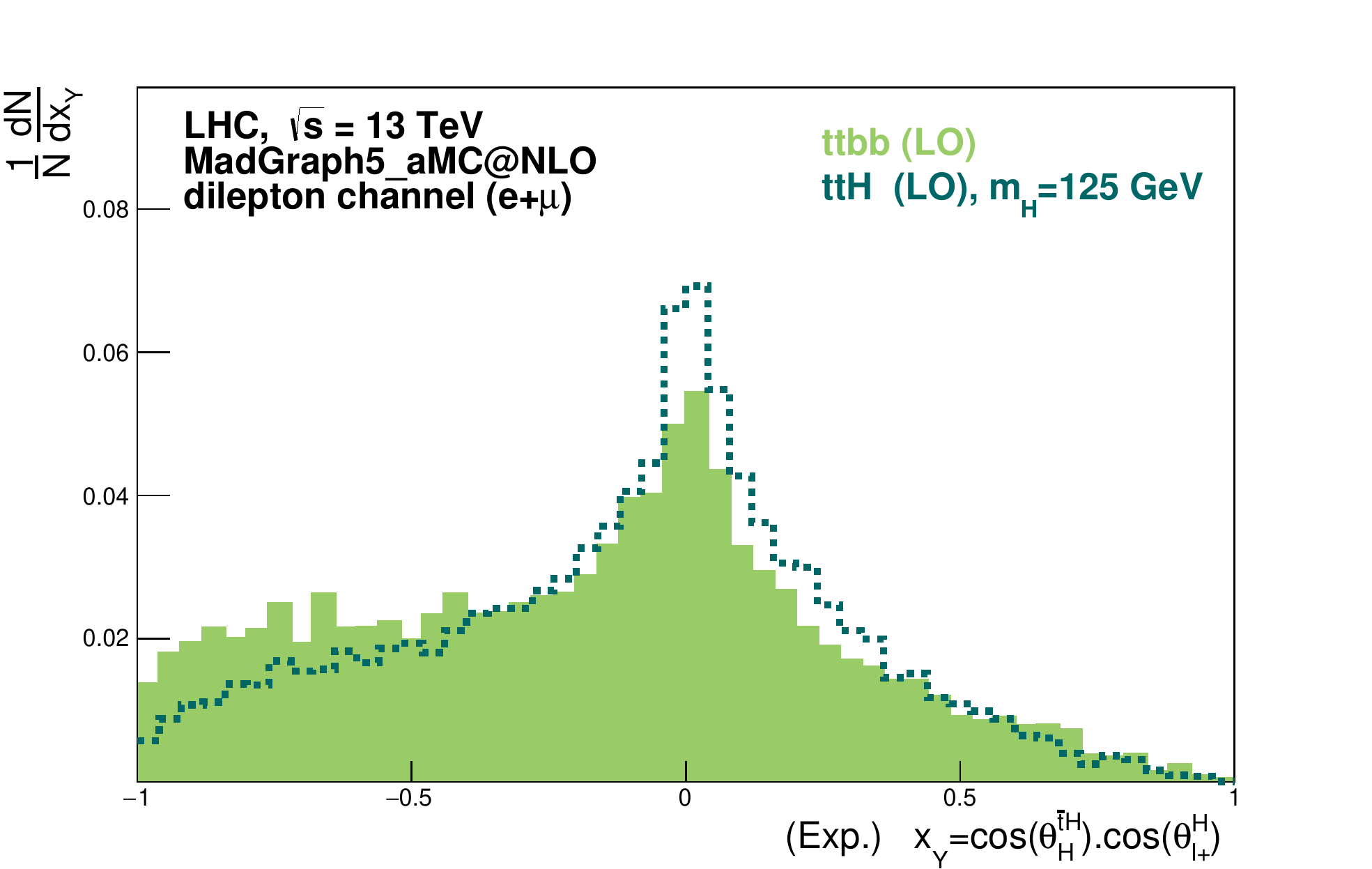,height=5.5cm,clip=} & \quad & 
\epsfig{file=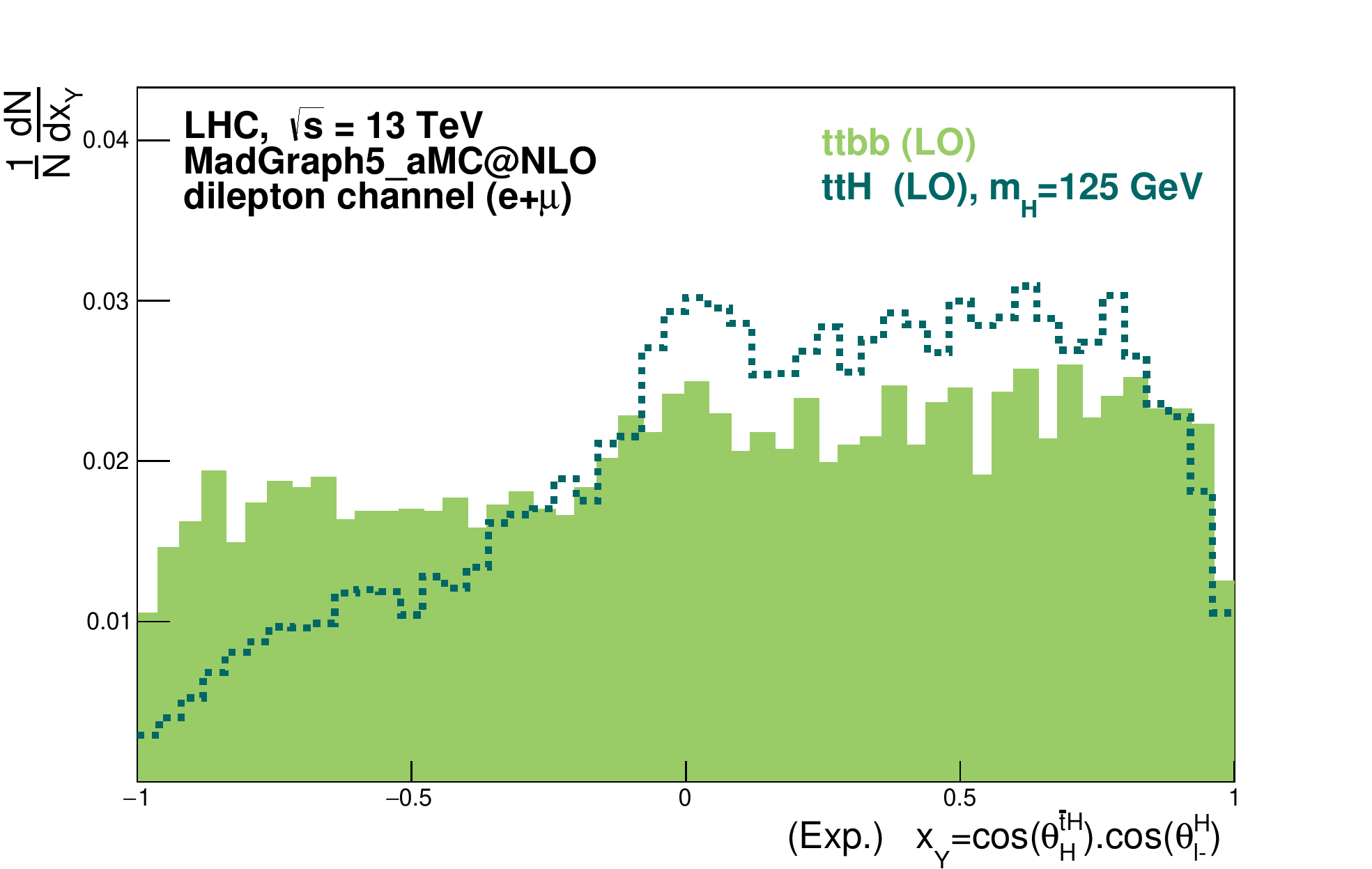,height=5.5cm,clip=}
\end{tabular}
\caption{Same as in figure.~\protect\ref{fig:RecAng1}, but without truth match.}
\label{fig:ExpAng1}
\end{center}
\end{figure*}

\begin{figure*}
\begin{center}
\begin{tabular}{ccc}
\epsfig{file=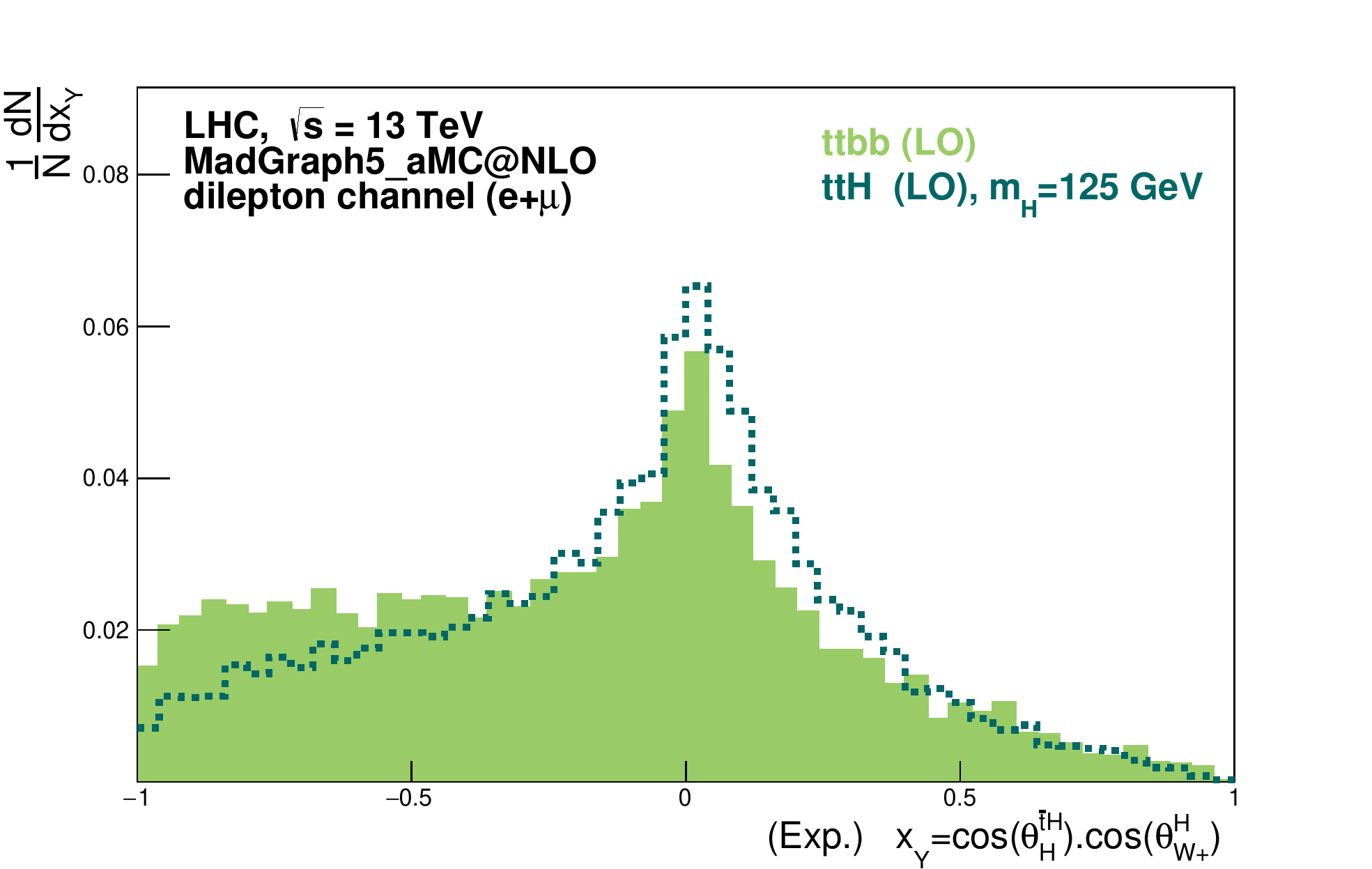,height=5.5cm,clip=} & \quad &
\epsfig{file=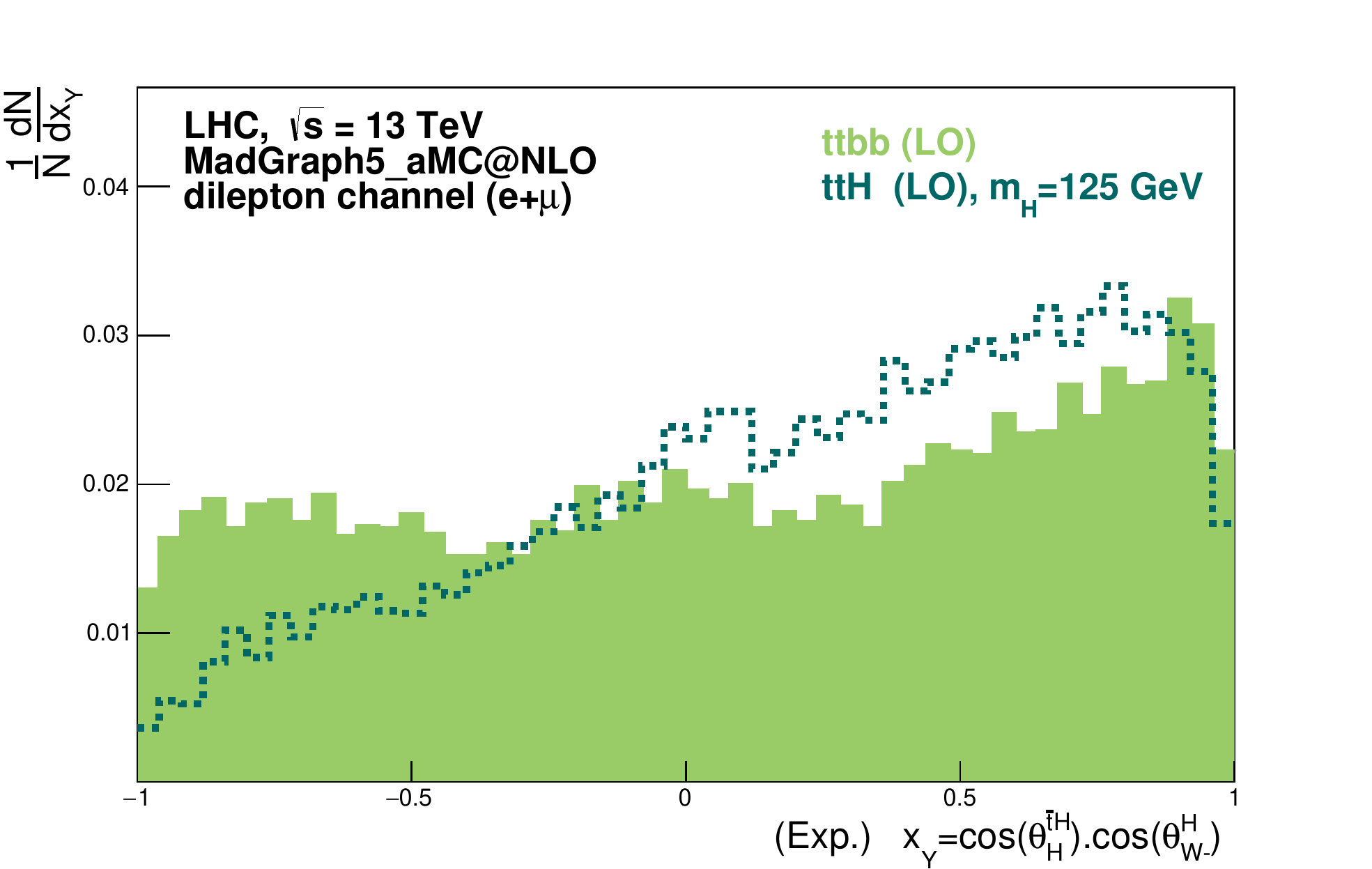,height=5.5cm,clip=}
\end{tabular}
\caption{Reconstructed product (without truth match) of the cosine of the angle between the Higgs momentum direction (in the $\bar{t}H$ centre-of-mass) with respect to the $\bar{t}H$ direction (in the $t\bar{t}H$ system),  and the cosine of the angle of the $W^+$(left) and $W^-$(right) momentum (in the Higgs centre-of-mass system) with respect to the Higgs direction (in the $\bar{t}H$ system).}
\label{fig:ExpAng2}
\end{center}
\end{figure*}

\begin{figure*}
\begin{center}
\begin{tabular}{ccc}
\epsfig{file=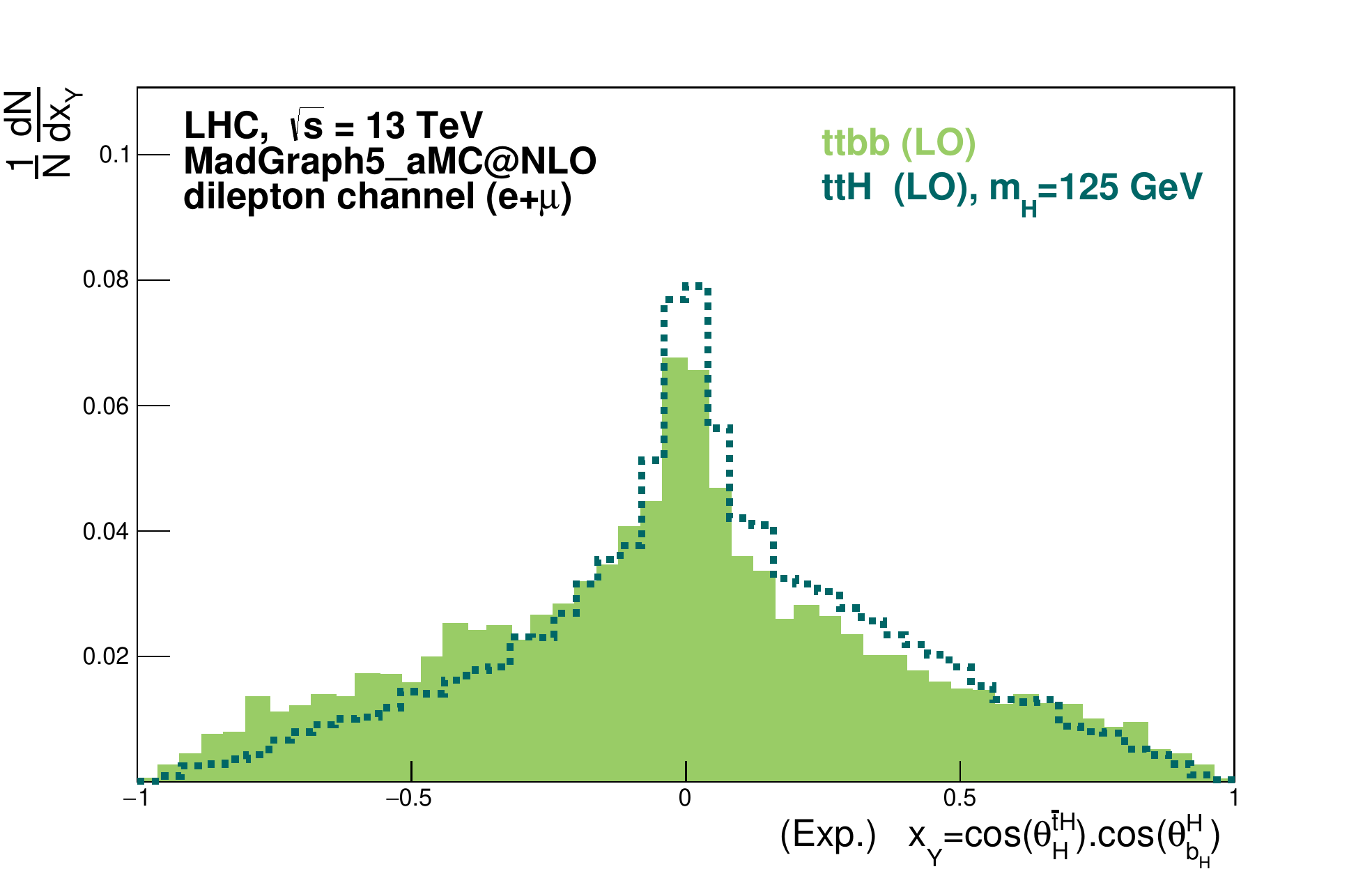,height=5.5cm,clip=} & \quad &
\epsfig{file=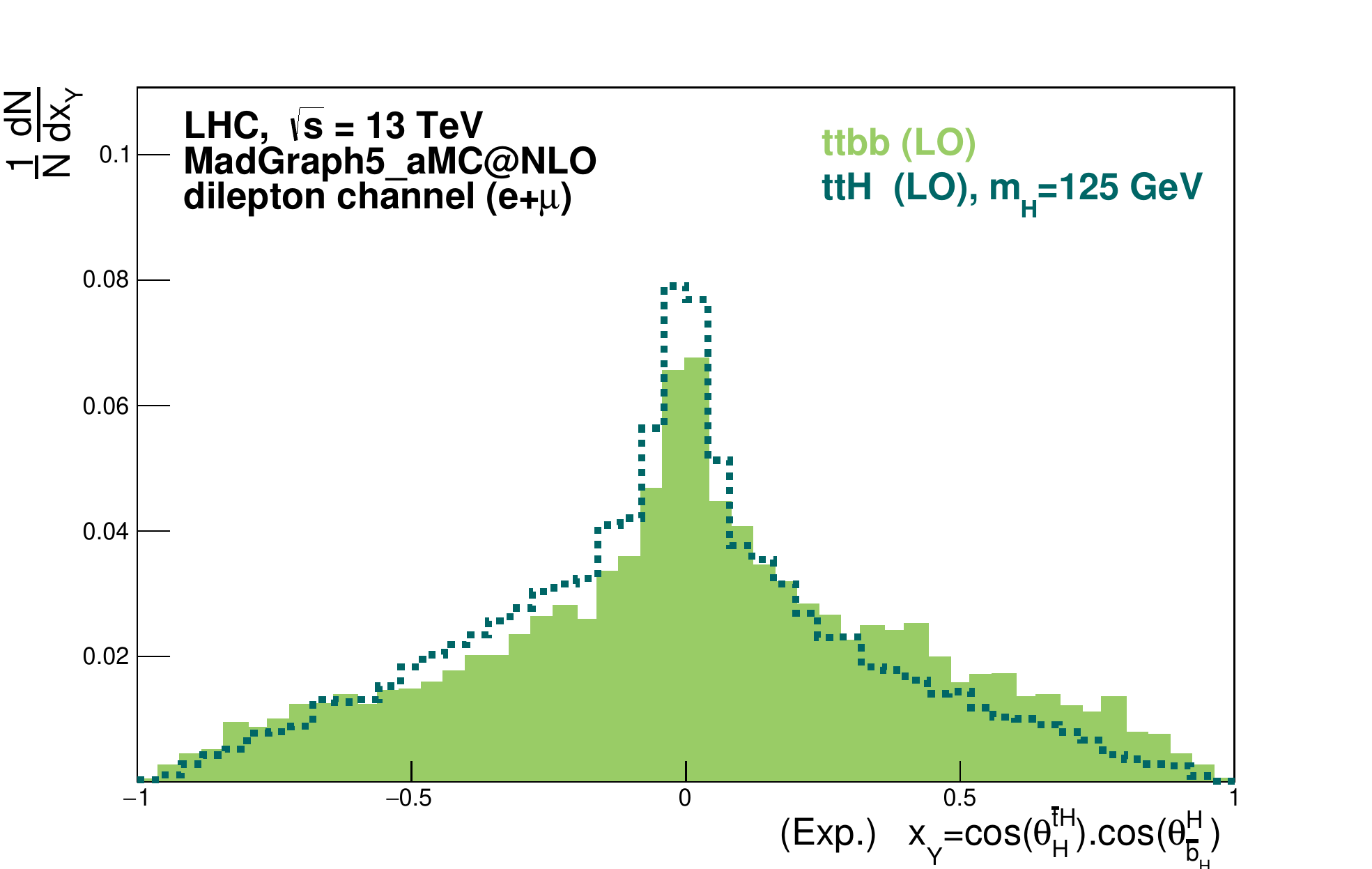,height=5.5cm,clip=}
\end{tabular}
\caption{Reconstructed product (without truth match) of the cosine of the angle between the Higgs momentum direction (in the $\bar{t}H$ centre-of-mass) with respect to the $\bar{t}H$ direction (in the $t\bar{t}H$ system),  and the cosine of the angle of the Higgs $b$-quark(left) and $\bar{t}$-quark(right) momentum (in the Higgs centre-of-mass system) with respect to the Higgs direction (in the $\bar{t}H$ system).}
\label{fig:ExpAng3}
\end{center}
\end{figure*}

In figure~\ref{fig:GenNCAng1} we show distributions at parton level,
without any cuts, for the product of $\cos{(\theta^{\bar{t} H}_{H})}$,
and $\cos{(\theta^{H}_{Y})}$, for $Y=\ell^+$(left) and
$Y=\ell^-$(right). We can see the distributions are quite different
between signal and background events. The effect of applying the $p_T$
and $\eta$ cuts to jets and leptons is seen in
figure~\ref{fig:GenWCAng1}. A clear reduction on the number of events
is observed due to the cuts applied. In figure~\ref{fig:RecAng1} we
can see the effect of the kinematic fit reconstruction, still with
the truth match information. The information on the
angular distribution is preserved to a large extent, even after the full kinematical
reconstruction. As we will see this is also true when the reconstruction is
performed without truth match. In figure~\ref{fig:ExpAng1} we show the
reconstructed product (without truth match) of $\cos{(\theta^{\bar{t}
    H}_{H})}$ and $\cos{(\theta^{H}_{Y})}$, for $Y=\ell^+$(left) and
$Y=\ell^-$(right). In figures~\ref{fig:ExpAng2} and ~\ref{fig:ExpAng3}
we show the same distributions but with the charged leptons replaced
by the $W$-bosons and $b$-quarks from the Higgs decay,
respectively. It is quite apparent that some of the angular
distributions allow discrimination between signal and
background even after the full kinematical reconstruction without
the truth match. Since we did not try to optimise the kinematical
reconstruction, it is foreseeable that better results could be
obtained in the future.

\section{Forward-Backward Asymmetries}

\begin{figure*}
\vspace*{5cm}
\begin{center}
\begin{tabular}{ccc}
\epsfig{file=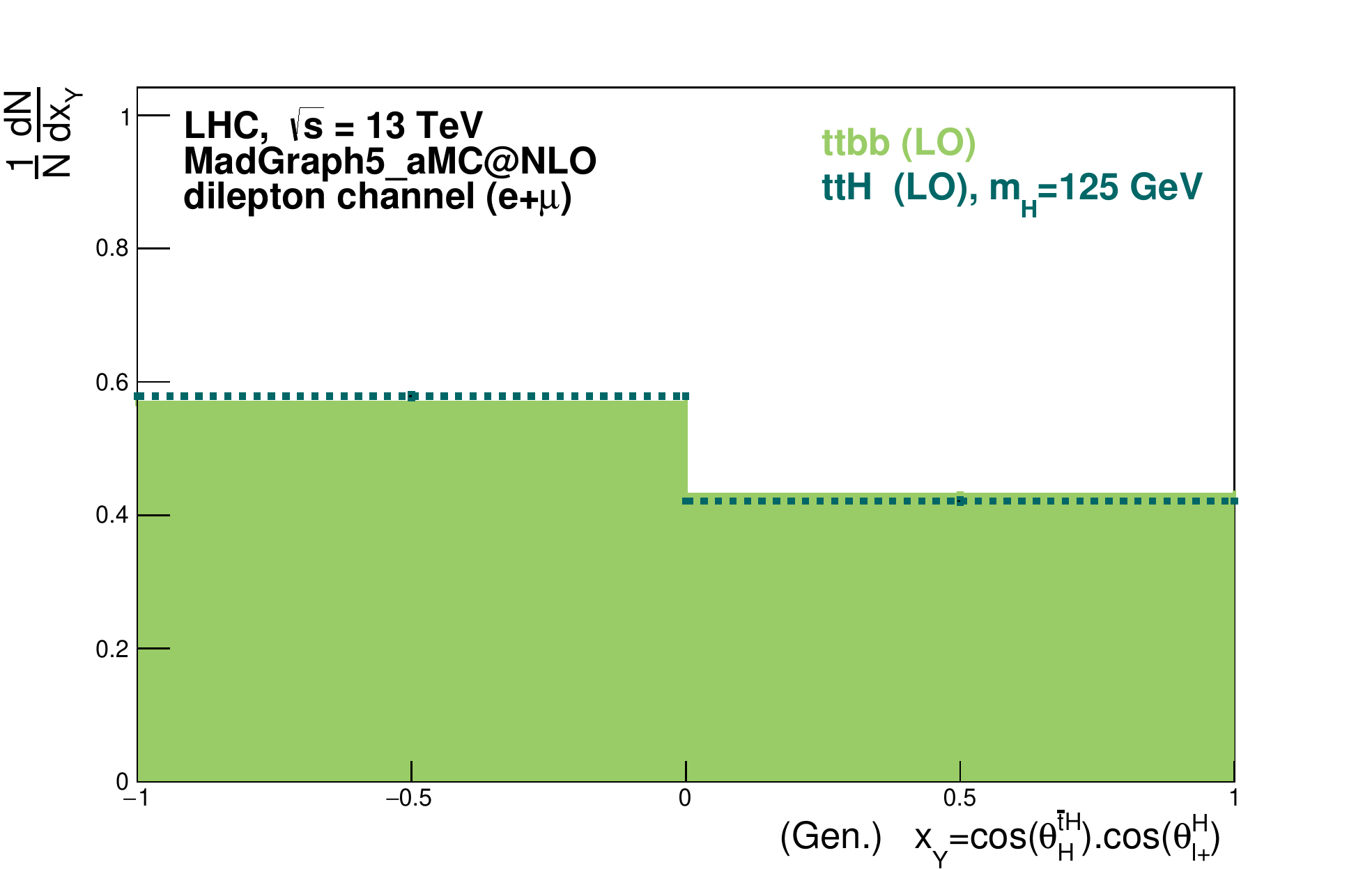,height=5.3cm,clip=} & \quad &
\epsfig{file=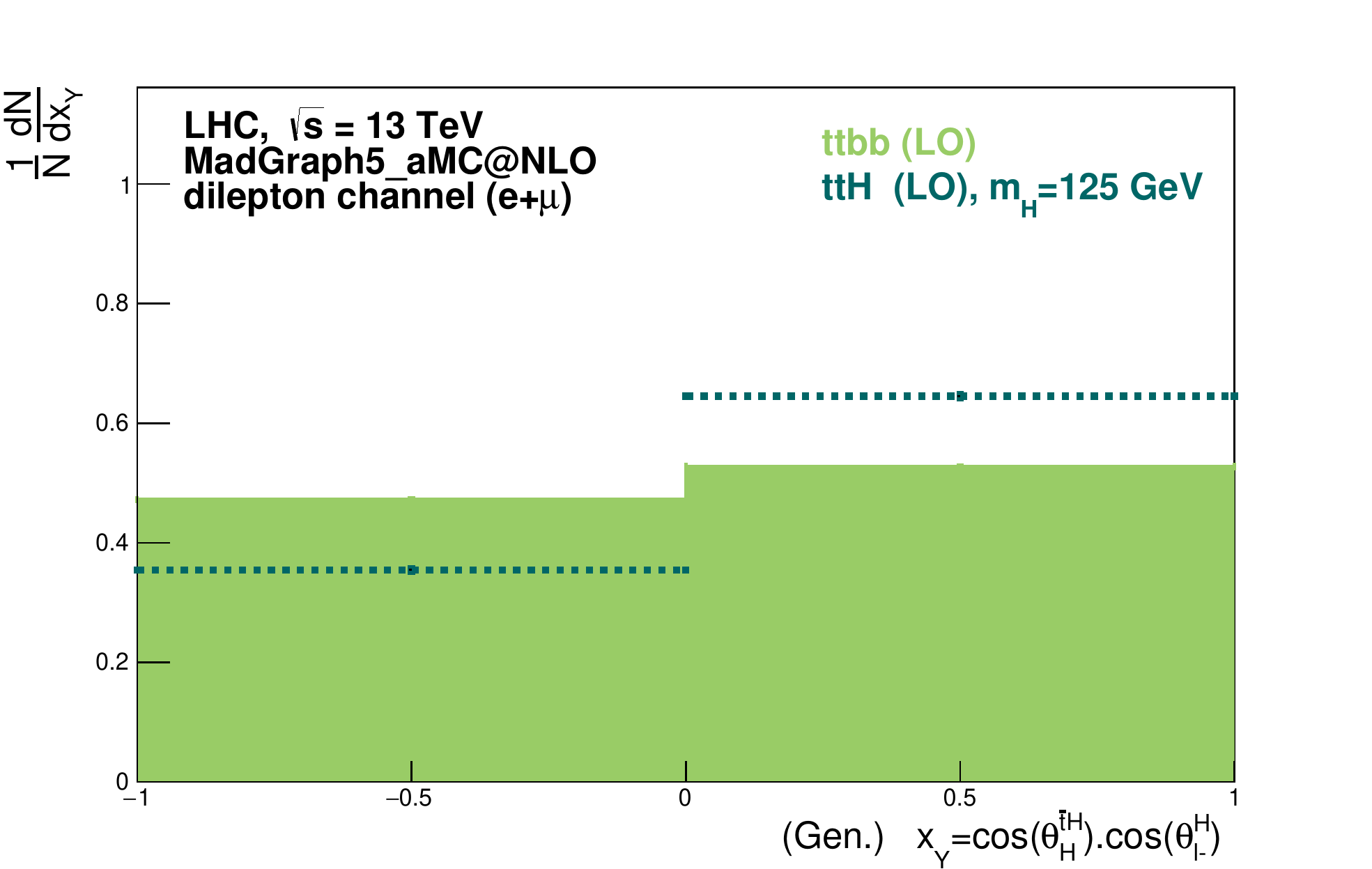,height=5.3cm,clip=}
\end{tabular}
\caption{Two binned generated product of the cosine of the angle between the Higgs momentum direction (in the $\bar{t}H$ centre-of-mass) with respect to the $\bar{t}H$ direction (in the $t\bar{t}H$ system),  and the cosine of the angle of the $\ell^+$(left) and $\ell^-$(right) momentum (in the Higgs centre-of-mass system) with respect to the Higgs direction (in the $\bar{t}H$ system).}
\label{fig:GenNCAsym1}
\end{center}
\end{figure*}

Based on the angular distributions introduced in the previous section, we propose to use several forward-backward 
asymmetries ($A^Y_{FB}$) in this paper, defined using the double angular product \\
\begin{eqnarray}
\ensuremath{x_Y= \cos{(\theta^{\bar{t} H}_{H})} \times \cos{(\theta^{H}_{Y})}}.
\label{equ:a8}
\end{eqnarray}
The asymmetries can be easily calculated, both at parton level and after the kinematic fit reconstruction, and are defined as, 
\begin{eqnarray}
\ensuremath{A^Y_{FB}= \frac  {    N(x_Y>0)-N(x_Y<0)   }{ N( x_Y>0)+N(x_Y<0) } },
\label{equ:a9}
\end{eqnarray}
where $N(x_Y>0)$ and $N(x_Y<0)$ correspond to the total number of events in the corresponding angular distribution with $x_Y$ above and below zero, respectively.
These asymmetries can be quite different between the signal $t\bar{t}H$ and the irreducible background $t\bar{t}b\bar{b}$. In Table~\ref{tab:AsymGenExp} we present the values of the asymmetries, with no cuts applied to the events,
 at parton level (and at LO) for different choices of the final state particle ($Y$) that is boosted to the centre of mass of the Higgs boson. As we can see, there are clear differences for some of the asymmetries i.e., $A^{Y=\ell- }_{FB}$, $A^{Y=W-  }_{FB}$, $A^{Y=\bar{b}}_{FB}$  ($\bar{b}$ from $\bar{t}$), between signal and background. We show in figure~\ref{fig:GenNCAsym1} an example of two-binned angular distributions for $Y=\ell^-$ and $Y=\ell^+$, respectively, evaluated at parton level without any $p_T$ or $\eta$ cuts applied to the events. 
\begin{table}[h]
\renewcommand{\arraystretch}{1.3}
\begin{center}
  \begin{tabular}{lccccc}
    \toprule
    (Asymmetries @ LO)                  		& \multicolumn{2}{c}{Parton level}              & & \multicolumn{2}{c}{Reconstruction} \\
    \toprule
                                                        & $t\bar{t}H$           &   $t\bar{t}b\bar{b}$  & & $t\bar{t}H$         & $t\bar{t}b\bar{b}$ \\
    \midrule
      $A^{Y=\ell+}_{FB}$   				&    	$-0.157$	&    $-0.137$  		& &   	$-0.141$	& $-0.268$  \\    
      $A^{Y=\ell- }_{FB}$    				&    	$+0.291$        &    $+0.056$   	& &   	$+0.331$	& $+0.118$  \\    
      $A^{Y=W+ }_{FB}$    				&    	$-0.154$	&    $-0.119$  		& &   	$-0.119$	& $-0.275$  \\    
      $A^{Y=W-  }_{FB}$    				&    	$+0.317$	&    $+0.067$  		& &   	$+0.348$	& $+0.127$  \\    
      $A^{Y=b}_{FB}$ ($b$ from $t$)    			&    	$-0.155$	&    $-0.141$  		& &   	$-0.179$	& $-0.306$  \\    
      $A^{Y=\bar{b}}_{FB}$ ($\bar{b}$ from $\bar{t}$)     &    	$+0.293$	&    $+0.053$  		& &   	$+0.334$	& $+0.117$  \\    
      $A^{Y=b}_{FB}$ ($b$ from $H$)		    	&    	$+0.000$	&    $+0.001$ 		& &   	$+0.086$	& $-0.048$  \\
      $A^{Y=\bar{b}}_{FB}$ ($\bar{b}$ from $H$) 		&    	$+0.000$	&    $-0.001$ 		& &   	$-0.086$	& $+0.048$  \\   
    \bottomrule
  \end{tabular}
\caption{{Values for the asymmetry for $t\bar{t}H$ and $t\bar{t}b\bar{b}$ events at the LHC. The second and third column show the observed asymmetries at the parton level (without any cuts), while the fourth and last column show same asymmetries after applying the selection cuts and the kinematical reconstruction (without truth match).}}
\label{tab:AsymGenExp}
\end{center}
\end{table}

\begin{figure*}
\begin{center}
\begin{tabular}{ccc}
\epsfig{file=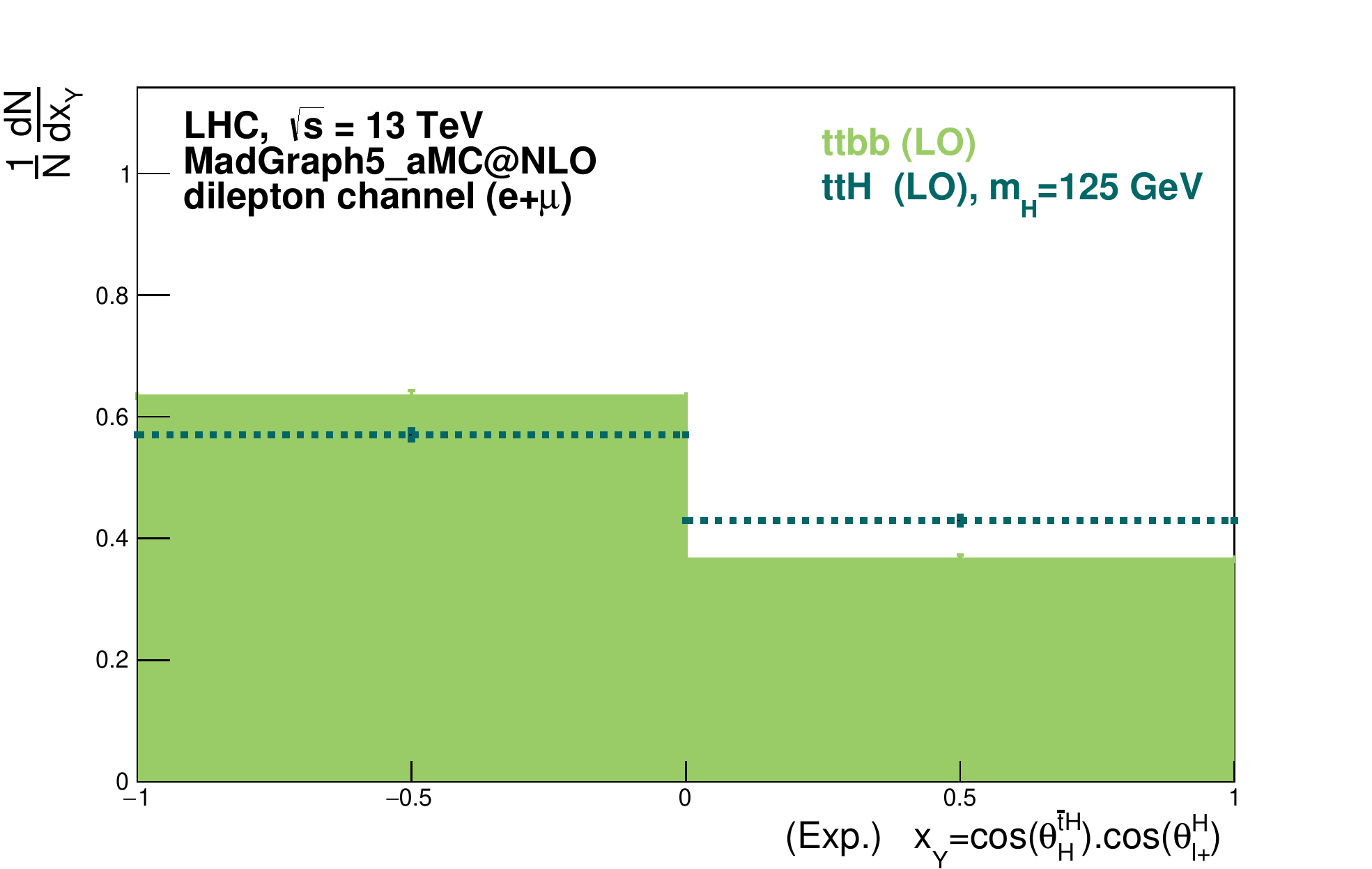,height=5.3cm,clip=} & \quad &
\epsfig{file=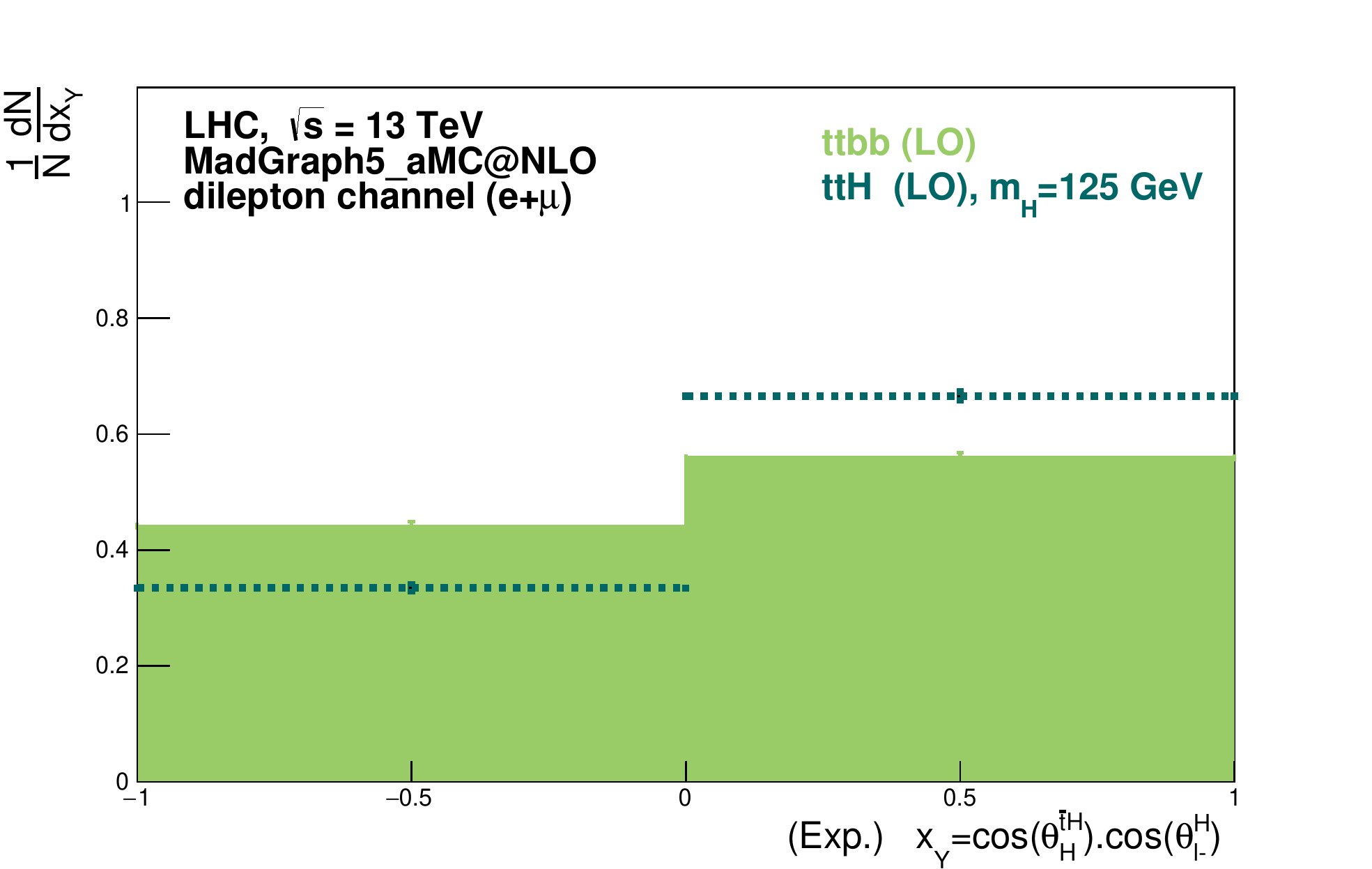,height=5.3cm,clip=}
\end{tabular}
\caption{Two binned reconstructed product (without truth match) of the cosine of the angle between the Higgs momentum direction (in the $\bar{t}H$ centre-of-mass) with respect to the $\bar{t}H$ direction (in the $t\bar{t}H$ system),  and the cosine of the angle of the $\ell^+$(left) and $\ell^-$(right) momentum (in the Higgs centre-of-mass system) with respect to the Higgs direction (in the $\bar{t}H$ system).}
\label{fig:ExpAsym1}
\end{center}
\end{figure*}

In Table~\ref{tab:AsymGenExp} we also show the values of the asymmetries after all cuts and the kinematic fit reconstruction (without truth match), for different choices of the final state particle ($Y$) boosted to the centre of mass of the Higgs boson. As we can see, even after the kinematical reconstruction there are clear differences for some of the asymmetries i.e., $A^{Y=\ell- }_{FB}$, $A^{Y=W-  }_{FB}$, $A^{Y=\bar{b}}_{FB}$  ($\bar{b}$ from $\bar{t}$), between signal and background. Note that the two asymmetries $A^{Y=b}_{FB}$ ($b$ from $H$) and $A^{Y=\bar{b}}_{FB}$ ($\bar{b}$ from $H$) are zero at the parton level, but non-zero at the reconstructed level due to a non-perfect reconstruction of the event. We show in figure~\ref{fig:ExpAsym1} the two-binned angular distributions for $Y=\ell^-$ and $Y=\ell^+$, respectively, obtained after all cuts and full kinematical reconstruction of events. Although distortions (that may be corrected for) are visible as a consequence of the cuts applied and kinematical reconstruction, some of the angular distributions and asymmetries show significant differences between the signal and dominant background, even after reconstruction (see Table~\ref{tab:AsymGenExp}).

\begin{figure*}
\vspace*{5cm}
\begin{center}
\begin{tabular}{ccc}
 \quad & \epsfig{file=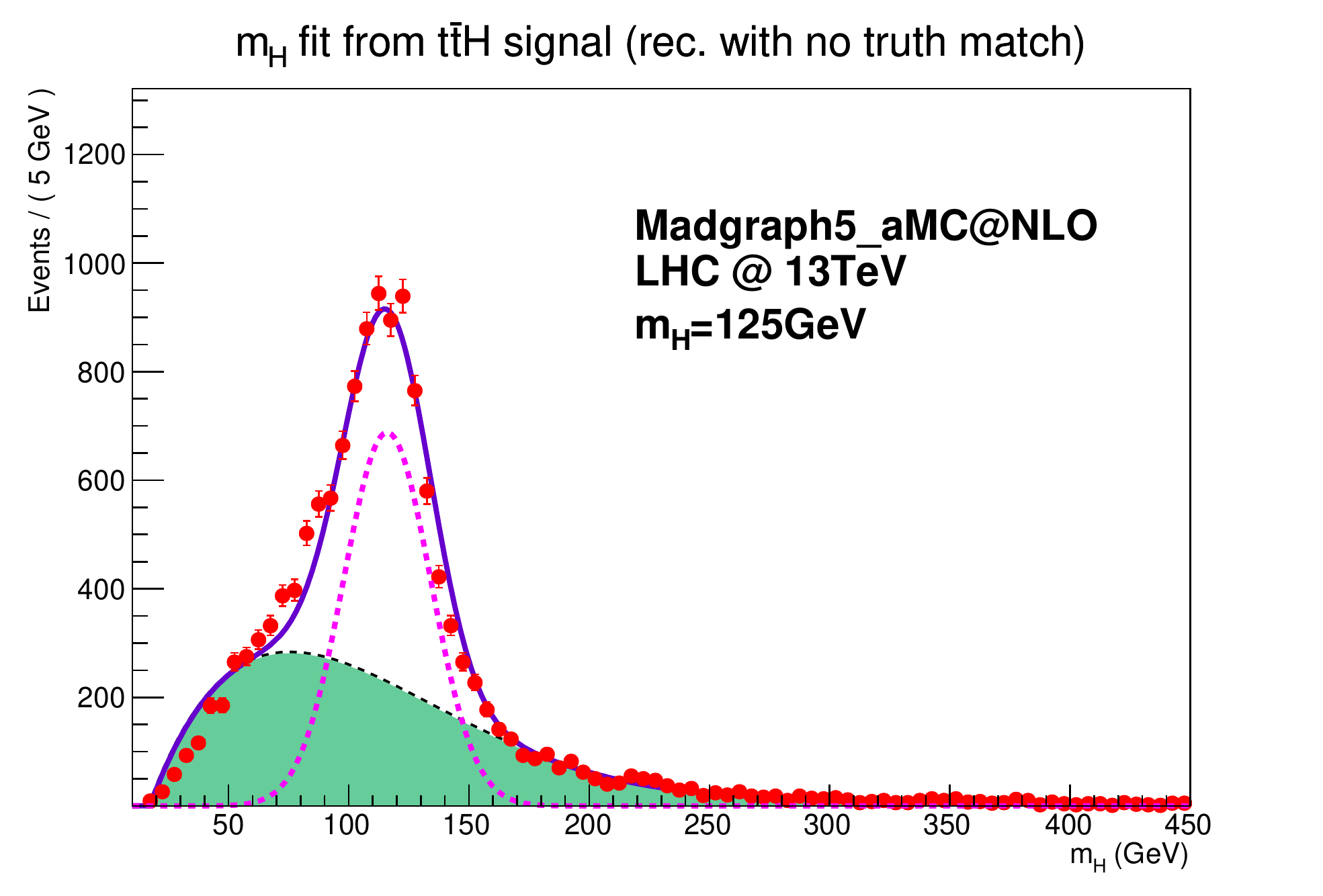,height=12.5cm,clip=} & 
\end{tabular}
\caption{Higgs reconstructed mass (without truth match) fit (see text for details). The dashed line correspond to the signal and shaded region the combinatorial background.}
\label{fig:ExpFit}
\end{center}
\end{figure*}

One last comment is in order in what concerns the reconstructed mass of the Higgs boson. Even after the full kinematical reconstruction and possible contamination from the combinatorial background arising whenever the reconstruction is performed without truth match, it is still possible to recognise, in the $m_{b\bar{b}}$ variable, the mass peak corresponding to the right combination of $b$-quarks coming from the Higgs boson. In figure~\ref{fig:ExpFit} we show a fit of the Higgs mass in signal events, just to guide the eye, performed with RooFit~\cite{Verkerke:2003ir} using a Chebychev polynomial (to parametrise the combinatorial background) and a Gaussian distribution (to describe the Higgs mass reconstructed from two $b$-quarks). Once again no optimisation is performed in the fit. The effect of the combinatorial background is clearly visible as a shoulder towards lower values of the invariant mass distribution which extends to higher values with a long continuous tail. We argue that it is important to understand the different components of the combinatorial background and dedicated studies must be performed to minimise the effect of its uncertainties, but this is largely outside the scope of this paper.\\

\section{Conclusions}

In this paper the $t{\bar t} H$ production in proton-proton collisions at the LHC is addressed, for a centre of mass energy of 13~TeV. Fully reconstructed, dileptonic final state  $t{\bar t}H$ events, from the decays $t\rightarrow bW^+ \rightarrow b \ell^+\nu_\ell$, $\bar{t}\rightarrow \bar{b}W^- \rightarrow \bar{b} \ell^-\bar{\nu}_\ell$ and $h\rightarrow b\bar{b}$, are used to probe new angular distributions and asymmetries that allow better discrimination between the signal and the main irreducible background. We show that it is possible to fully reconstruct $t\bar{t}H$ final states in the dileptonic topology and, even with a reconstruction which is not optimised, still be sensitive to the new angular distributions and asymmetries, which seem to be quite different between the signal and background even after full reconstruction.

One should again stress that current experimental results on the $pp \to t \bar t H$ are already very impressive
even though, essentially, kinematic variables are used. We have shown that the use of new variables 
that make use of the spin information of signal and background processes can further improve 
the results for the cross section measurement. Furthermore, the spin information that is present in the matrix elements
survives showering, detector simulation, selection and reconstruction, even in the most challenging decay channel of dileptonic $t\bar{t}H$ events.

\vspace{0.5cm}

\section*{Acknowledgements}

This work was partially supported by Funda\c{c}\~ao para a Ci\^encia e
Tecnologia, FCT (projects CERN/FP/123619/2011 and
EXPL/FIS-NUC/1705/2013, grants SFRH/BI/52524/2014 and
SFRH/BD/73438/2010, and contracts IF/00050/2013 and
IF/00014/2012). The work of M.C.N.~Fiolhais was supported by
LIP-Laborat\'orio de Instrumenta\c c\~ao e F\'isica Experimental de
Part\'iculas, Portugal (grant PestIC/FIS/LA007/2013).  The work of
R.S. is supported in part by FCT under contract
PTDC/FIS/117951/2010. Special thanks go to Juan Antonio
Aguilar-Saavedra for all the fruitful discussions and a long term
collaboration.


%
\vspace*{5cm}

\clearpage


%
%
%

\end{document}